\documentclass[12pt, letterpaper]{aastex631}
\usepackage[normalem]{ulem}
\useunder{\uline}{\ul}{}
\usepackage{graphicx}

\usepackage{floatrow}
\usepackage{gensymb}
\newfloatcommand{capbtabbox}{table}[][\FBwidth]
\floatstyle{plaintop}
\restylefloat{table} 
\usepackage{amsmath}
\usepackage{bm}
\newcolumntype{?}{!{\vrule width 1.5pt}}
\usepackage{tablefootnote}
\usepackage{pdfpages}
\makeatletter
\AtBeginDocument{\let\LS@rot\@undefined}
\makeatother

\definecolor{LightBlue}{rgb}{0.68, 0.85, 0.9}
\definecolor{LightCyan}{rgb}{0.88,1,1}
\definecolor{LightApricot}{rgb}{0.99, 0.84, 0.69}

\definecolor{Orange}{rgb}{0.902, 0.624, 0.00}
\definecolor{SkyBlue}{rgb}{0.337, 0.706, 0.914}
\definecolor{BluishGreen}{rgb}{0.00, 0.620, 0.451}
\definecolor{Yellow}{rgb}{0.941, 0.894, 0.259}
\definecolor{Blue}{rgb}{0.00, 0.447, 0.698}
\definecolor{Vermillion}{rgb}{0.835, 0.369, 0.00}
\definecolor{ReddishPurple}{rgb}{0.80, 0.475, 0.655}

\definecolor{SkyBlue}{rgb}{0.835, 0.369, 0.00}

\definecolor{Vermillion}{rgb}{0.337, 0.706, 0.914}

\graphicspath{{./}{figures/}}

\shorttitle{PSG Reflected Planetary Spectra}
\shortauthors{Saxena, Villanueva, Mandell, Zimmerman and Smith}

\begin{document}

\title{Simulating Reflected Light Exoplanet Spectra of the Promising Direct Imaging Target, $\upsilon$ And d, with a New, Fast Sampling Method using the Planetary Spectrum Generator}

\correspondingauthor{Prabal Saxena}
\email{prabal.saxena@nasa.gov}

\author{Prabal Saxena}
\affiliation{CRESST II/University of Maryland, College Park, Maryland 20742, USA}
\affiliation{NASA Goddard Space Flight Center, Greenbelt, Maryland 20771, USA}

\author[0000-0002-2662-5776]{Geronimo L. Villanueva}
\affiliation{NASA Goddard Space Flight Center, Greenbelt, Maryland 20771, USA}

\author[0000-0001-5484-1516]{Neil T. Zimmerman}
\affiliation{NASA Goddard Space Flight Center, Greenbelt, Maryland 20771, USA}

\author{Avi M. Mandell}
\affiliation{NASA Goddard Space Flight Center, Greenbelt, Maryland 20771, USA}

\author{Adam J. R. W. Smith}
\affiliation{Southeastern Universities Research Association, 1201 New York Ave., NW, Suite 430, Washington, DC 20005}
\affiliation{Center for Research and Exploration in Space Science and Technology, NASA Goddard Space Flight Center, Greenbelt, MD 20771}


\begin{abstract}

Simulations of exoplanet albedo profiles are key to planning and interpreting future direct imaging observations. In this paper we demonstrate the use of the Planetary Spectrum Generator (PSG) to produce simulations of reflected light exoplanet spectra.  We use PSG to examine multiple issues relevant to all models of directly imaged exoplanet spectra and to produce sample spectra of the bright, nearby exoplanet $\upsilon$ Andromedae d, a potential direct imaging target for next-generation facilities.  We introduce a new, fast, and accurate sub-sampling technique that enables calculations of disk-integrated spectra one order of magnitude faster than Chebyshev-Gauss sampling for moderate- to high-resolution sampling.  Using this method and a first-principles-derived atmosphere for $\upsilon$ And d, we simulate phase-dependent spectra for a variety of different potential atmospheric configurations. The simulated spectra for $\upsilon$ And d include versions with different haze and cloud properties. Based on our combined analysis of this planet's orbital parameters, phase- and illumination-appropriate model spectra, and realistic instrument noise parameters, we find that $\upsilon$ And d is a potentially favorable direct imaging and spectroscopy target for the Coronagraph Instrument (CGI) on the Nancy Grace Roman Space Telescope.  When a noise model corresponding to the Roman CGI SPC spectroscopy mode is included, PSG predicts the time required to reach a signal-to-noise ratio of 10 of the simulated spectra in both the central wavelength bin of the Roman CGI SPC spectroscopy mode (R=50 spectrum) and of the Band 1 HLC imaging mode is approximately 400 and and less than 40 hours, respectively. We also discuss potential pathways to extricating information about the planet and its atmosphere with future observations and find that Roman observations may be able to bound the interior temperature of the planet.   

\end{abstract}



\section{Introduction} \label{sec:intro}

The characterization of atmospheres and surfaces of exoplanets has become a frontier in understanding worlds outside our solar system.  Interpretation of exoplanet spectroscopy, particularly of transmission spectra, has been one of the most important tools for this characterization.  It has enabled the identification of specific atmospheric constituents, the detection of clouds and hazes in atmospheres and the inference of the atmospheric temperature and pressure structure on exoplanets \citep{2008ApJ...673L..87R, 2014Natur.505...66K, 2016Natur.537...69D}.  


However, transmission spectroscopy is primarily effective at probing the atmospheres of the close-in planets, since the probability of detection of a transit \citep{2013PASP..125..933S} and the frequency of transits drops rapidly with orbital distance. Observations of reflected light spectra provide a means of probing atmospheres of planets at larger orbital separations, enabling characterization in a complementary phase space to the types of planets that transmission spectroscopy has been most sensitive to. 

Observations of directly imaged exoplanets have been the focus of a number of efforts from both ground and space that are attempting to open the characterization of these worlds. Ground-based projects such as the Gemini Planet Imager \citep{2014PNAS..11112661M} and Spectro-Polarimetic High contrast imager for Exoplanets REsearch (SPHERE) \citep{2019A&A...631A.155B}, and space-based efforts such as the James Webb Space Telescope coronagraphs \citep{Krist2007SPIE, 2015PASP..127..633B} represent the present-day capabilities in this area. To date, all direct imaging detections have been limited to young, self-luminous, giant exoplanets; fainter reflected light signatures generally fall below the detection floors of existing instruments~\citep{Guyon2005ApJ}. The Coronagraph Instrument (CGI) on the Nancy Grace Roman Space Telescope (formerly the Wide Field Infrared Space Telescope, WFIRST), slated for launch in 2026, will be among the first to reach the $\sim10^{-8}$ planet-to-star contrast sensitivity needed to image exoplanets in reflected starlight~\citep{Traub2016JATIS, Mennesson2020arXiv}. Various observatory and instrument concepts have been proposed for the coming decades that could push direct imaging techniques to the contrast levels required to detect and characterize planets smaller than gas giants in visible and infrared wavelengths~\citep{2015IJAsB..14..279Q, 2006SPIE.6267E..2AA, 2018SPIE10702E..A5S, 10.1117/12.2240457, 2019arXiv191206219T}. 

These instrument development efforts have been motivated by a significant body of literature that has attempted to model spectra of planets that may be observed using direct imaging.  Studies have examined potential exoplanet spectra over a wide range of phase space with variation due to both planetary parameters and orbital system parameters \citep{2000ApJ...538..885S, 2004ApJ...609..407B, 2010ApJ...724..189C}.  A number of targeted studies have also examined the influence on reflection spectra of specific topics exploring variation in planetary parameters, atmospheric chemistry, and from noise models.  Some research has also attempted to simulate future observations with observatories such as the Roman Space Telescope \citep{2017SPIE10400E..0BR, 2019AJ....157..132L} given the planned mission and instrument parameters \citep{2017SPIE10400E..1PS, 2018SPIE10698E..2IM}. Finally, interpretation of potential future data given the context of existing models and studies have led to the exploration of retrieval schemes \citep{2016AJ....152..217L} and packages that may be used to simulate radiative transfer \citep{2019ApJ...878...70B} or carry out basic retrievals for targeted parameters from reflected spectra of certain exoplanets \citep{2020AJ....159..175D}. 

\begin{figure*}[b]
  \centering
  \includegraphics[scale=0.70]{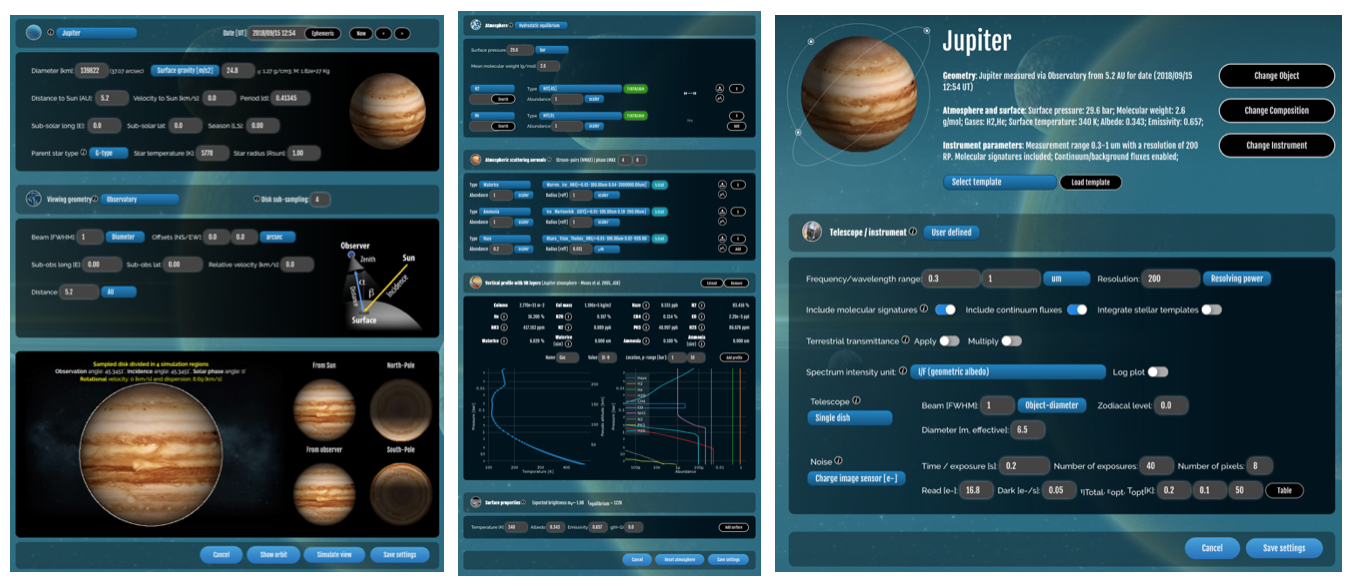}
  \caption{PSG modules for the Target and Geometry, Atmosphere and Surface, and Instrument Parameters from left to right.  The modules enable customization of a simulation from the website.  Choices in the modules displayed in this figure correspond to Jupiter spectra validation simulations with a haze in section~\ref{subsec:validation} .}
  \label{fig:KarkSetup}
\end{figure*}

In this paper, we describe the capabilities of the Planetary Spectrum Generator (PSG, https://psg.gsfc.nasa.gov, \textcite{2018JQSRT.217...86V}) to simulate reflected light spectra for directly imaged exoplanets.  In Section~\ref{sec:sampling}, we detail a newly developed disk sub-sampling technique we have implemented in PSG that enables accurate disk sampling required to produce reflected light spectra simulations with significantly less computational expense than current popular methods.  We also briefly describe the radiative transfer capabilities of PSG and the different customizable user friendly modules available that allow the user to modify the observing geometry and orbital properties, the atmosphere and surface properties and the instrumental parameters for a particular simulation.   Section~\ref{subsec:validation} discusses our validation of PSG's simulation capabilities by demonstrating the ability to reproduce Jupiter's spectra given appropriate input atmospheric profiles and then discusses comparison of PSG's modeling of Jupiter-like exoplanets to a previous study \citep{2018ApJ...858...69M}.  Section~\ref{sec:opacitytables} discusses the influence and effects of using different opacity values in simulating exoplanet reflection spectra.  Section~\ref{sec:UpsAndd} simulates spectra of $\upsilon$ And d using PSG.  $\upsilon$ And d is a bright, relatively high signal to noise direct imaging exoplanet target (considered one of the detectable RV planets by \textcite{2015arXiv150303757S}) and we consider its unique orbital and potential planetary parameters to explore the range of spectra that may be observed depending on those properties.  We further convolve this with the Roman Space Telescope's latest observational capabilities to produce mission-relevant simulated spectra. Finally, in Section~\ref{sec:Discussion} we discuss our results, future work and open questions.  


\section{Disk Sampling Methods} \label{sec:sampling}

Computation of accurate reflected light spectra of a spherical body, and specifically in this paper, reflected light spectra of directly imaged exoplanets, requires appropriate disk integration of the flux from the planet.  Using the same definition of albedo spectra as in previous works regarding reflected light spectra of exoplanets \citep{2010ApJ...724..189C}, we examine means of accurately integrating over
the emergent intensities from the planet that account for the various angles of incidence and emission for different portions of the planet. Correctly accounting for the emergent flux from the projected disk of the planet is critical in order to accurately integrate over portions of the planet closer to limb, particularly as observations are made at various phase.  This is because observations of the three dimensional planet appear flattened in the two dimensional plane of the sky due to a projection effect, which results in an emergent flux that appears to come from a projected two dimensional disk of the planet.  Sampling the apparent disk isotropically in a manner that takes into account the projection effect is key to calculating accurate emergent flux. 

\subsection{Exoplanet Reflection Spectra Using PSG} \label{sec:ExoPSG}

We use the the Planetary Spectrum Generator to examine these sampling techniques and to calculate the spectra in this paper.  The Planetary Spectrum Generator is a flexible radiative transfer suite that allows users to implement targeted observing scenarios through the integration of a range of spectroscopic,  atmospheric and instrument databases.  PSG enables users to synthesize a broad range of spectra through a user-friendly web interface to these models and  databases.  The full description of the tool is beyond the scope of this paper, but an overview of PSG is given in \textcite{2018JQSRT.217...86V}.  Different modules allow the user to specify a particular scene, either directly through the web interface or though the application program interface, which enables queued runs of multiple simulations. This is possible by creating scripts that can modify configuration (labelled `config' from now on) files that PSG uses as input for a simulation.  Each simulation run on the online interface of PSG also allows a user to download the config file associated with the simulation. We provide several config files in the appendix as a reference for the reader. Users have the ability to customize a number of different parameters in a simulation, with modules that enable customization of orbital properties and observational geometry, surface and atmosphere properties and instrument properties. As a visual example, Figure~\ref{fig:KarkSetup} displays the module set up that was used (in addition to simulations using scripts that called the API) in the Jupiter spectra validation simulations in section~\ref{subsec:validation}.

Users may also upload settings for specific observational targets using pre-loaded templates or a lookup function where PSG extracts orbital parameters from the NASA Exoplanet Archive \citep{2013PASP..125..989A} for known exoplanets. The atmosphere and surface module also contains a number of preset atmospheric templates the user can choose, including many specifically designed for or relevant to exoplanets. Two examples are a set of templates that compute temperature/pressure profiles \citep{2014A&A...562A.133P} and line by line abundances \citep{Kempton_2017} for gas giant exoplanets with varying chemistry, and another set that produces templates for terrestrial exoplanets \citep{2017EPSC...11..298T}.


\begin{figure*}
  \centering
  \includegraphics[scale=0.75]{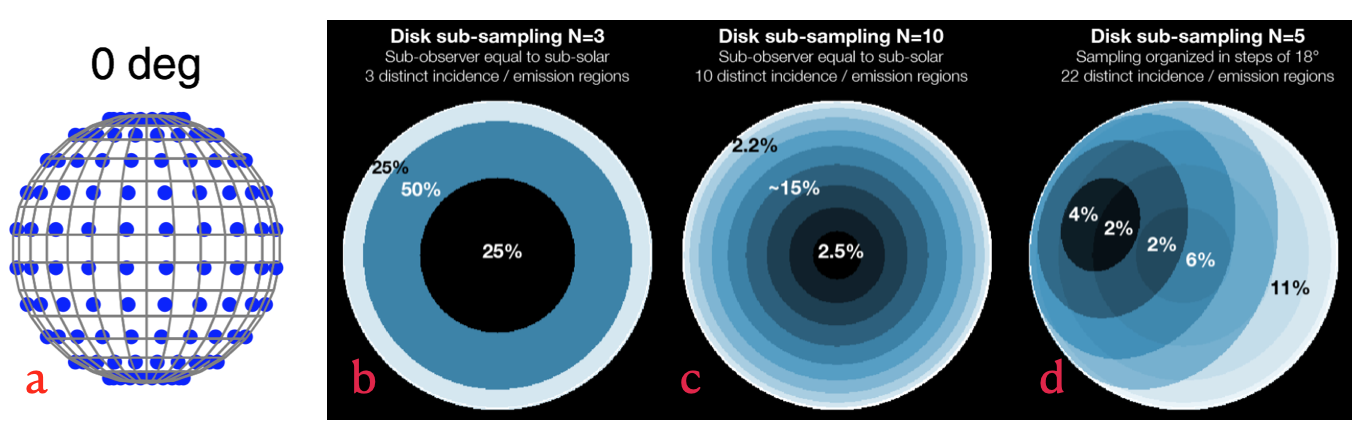}
  \caption{Visual display of sampling methods used in this paper.  The sphere on the left (a) is sampled with 10 by 10 Chebyshev-Gauss points, which can be used to calculate the (visible) disk integrated flux.  The right panel corresponds to the disk subsampling method.  Sampling regions (b, c, and d) are displayed for different subsample choices for an observation vantage equal to sub-solar for b and c and for another case, d, where incidence and emission angle are not coincident due to a symmetry.  Note sub-solar examples are for demonstration and comparison purposes relative to Chebyshev-Gauss sampling, as an exactly sub-solar view geometry would be occulted by the host star.  While near sub-solar geometry would statistically be relatively unlikely at any given time, observations are likely to be optimized for such a geometry given the favorable signal from predominantly dayside reflected light.  }
  \label{fig:samplingvisual}
\end{figure*}

\subsection{Sampling Techniques: Chebyshev-Gauss Integration} \label{subsec:chebgauss}

A common method of integrating the flux for these planets involves sampling the emergent flux from a sphere using many plane--parallel facets, where facets correspond to different pairs of incidence and emergence angles.  This has been used in numerous studies \citep{2010ApJ...724..189C, 2015ApJ...804...94W, 2018ApJ...858...69M} examining reflected light spectra of directly imaged exoplanets. In these and many other studies, the method has relied upon the ability to formulate the planetary albedo in a manner that can employ common numerical integration techniques to solve the integral for the flux. In this case, Chebyshev--Gauss quadrature using Chebyshev polynomials of the second kind are used to yield an exact result for polynomials of degree 2n - 1 or less by choosing suitable Chebyshev and Gaussian nodes/angles and weights at which to evaluate the flux \citep{1950ApJ...112..445H, 1965ApJS...11..373H}.  These are available online for any given number of nodes or can be calculated using formulas given in \textcite{2015ApJ...804...94W}.  Precision to the real value can be improved by increasing the number of nodes at which flux is evaluated, and numerical techniques for improving precision are also available \citep{DEHGHAN2005431}.  A visual example of the distribution of nodes using Chebyshev-Gauss integration using a 10x10 (100 total) set of nodes on a sphere at full phase is given in Figure~\ref{fig:samplingvisual}a. 

\begin{figure*}
  \centering
  \includegraphics[scale=0.50]{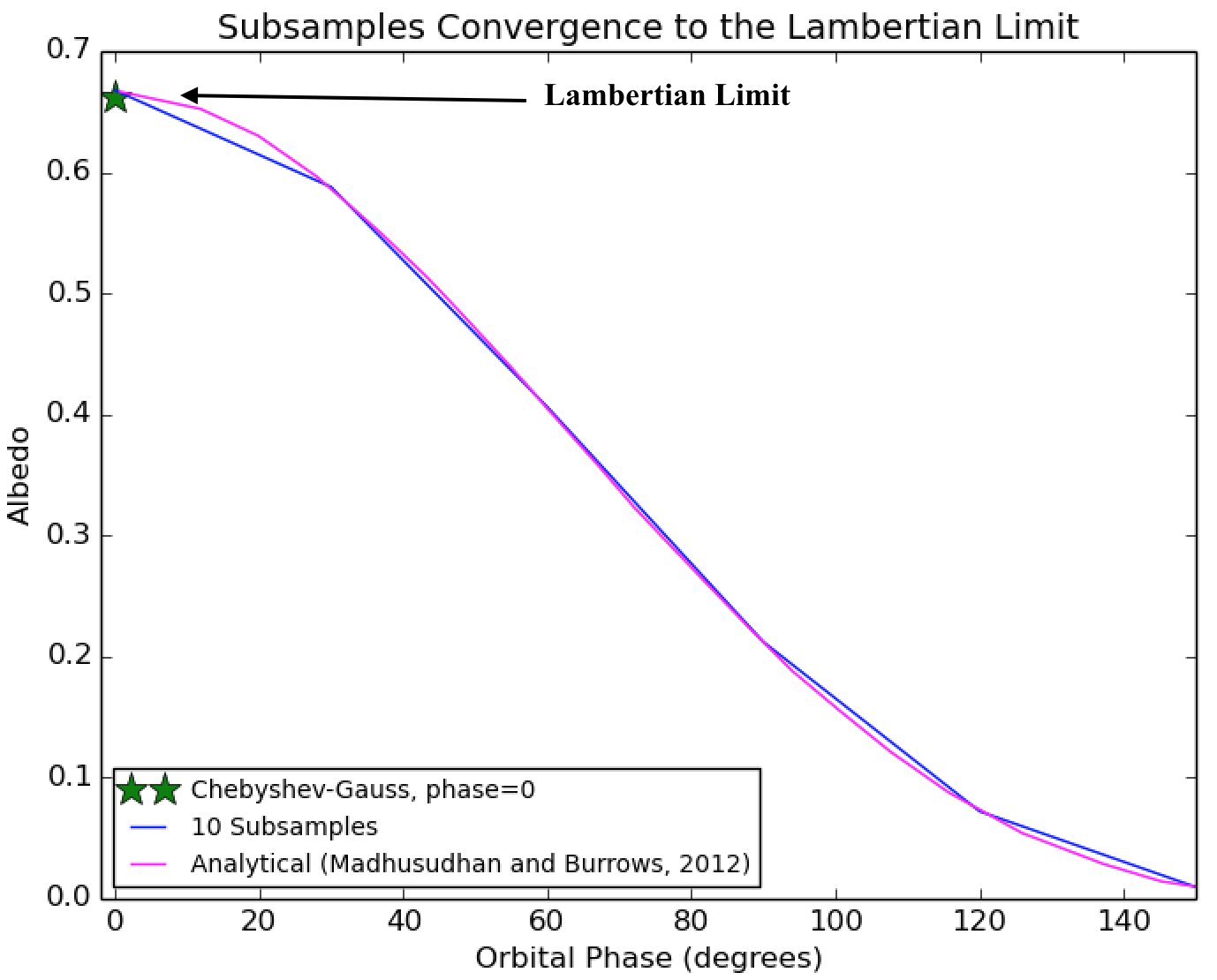}
  \caption{A plot of the PSG sub-sample method compared to the analytic Lambert scattering phase function \citep{2012ApJ...747...25M}.  There is strong agreement between the two ($<$1\% difference with sufficient samples - the 10 sample case is shown here), and both the sub-sample and Chebyshev-Gauss method as implemented in PSG converge to the Lambertian limit of 2/3 at full phase.}
  \label{fig:lambertconvergence}
\end{figure*}

Chebyshev--Gauss integration is easily implementable in PSG by calling the API and was used in order to calculate reflected light spectra and validate accuracy and efficiency of the subsampling method described in Section~\ref{subsec:subsample}.  Since simulations use a finite beam size (for example, see the annular beam size in white around Jupiter in the left portion of figure \ref{fig:KarkSetup}), the integrated flux calculated using Chebyshev--Gauss integration needs to be normalized by the beam size, which with a circular beam is just the area of the spherical cap.  In addition, Chebyshev-Gauss points across longitude need to be normalized to their limb-to-limb chord length, which is dependent on the latitude of interest and is largest at the equator.  In order to validate that Chebyshev--Gauss integration was being implemented correctly, we calculated fluxes for different scenes in order to compare to theoretical limits and to other data and simulations. We validated our implementation of the Chebyshev--Gauss integration with PSG/API by running both it and the subsampling method we developed, for a perfect reflecting surface (albedo=1) and iterating over the number of sampling points. We compared the albedo using our sampling method to the analytic Lambert scattering phase function as given in \textcite{2012ApJ...747...25M} and compared all the values at zero phase, where all three methods converged to the theoretical 2/3 limit as the number of sampling points increased (see figure \ref{fig:lambertconvergence}). 

\subsection{Sampling Techniques: A New Fast and Accurate Sub-sampling Technique} \label{subsec:subsample}

 In order to enable a more rapid but similarly accurate integration of disk-integrated flux, we develop a numerical algorithm that creates a 2D matrix of regions of similar outgoing flux based on the stellar incidence angle and observer-atmosphere angle. The disk is then divided across this 2D array by selecting the number of sub-samples that are used to encapsulate regions and define contributions/weighting functions for each eigenvalue. Internally, PSG divides the sampled disk into a map of 140 x 140 pixels, and computes incidence and emission angles for each pixel (19,600 sets). Without sub-sampling, PSG computes a single set of effective incidence and emission angle from these 19,6000 sets, which is then used to compute a single radiative-transfer calculation. When sub-sampling is enabled, PSG creates histograms of incidence/emission angles, with N defining the number of bins between 0 and 90 degrees. It then identifies the different possible combinations, and the relative occurrence of each incidence/emission combination (among the 19,600 sets), ultimately establishing the weight for those super-sets. PSG then computes radiative-transfer calculations using those super-sets and adds the spectra for each case yet weighted by the numerically computed weights. For instance, for a symmetric simple case (sub-observer angle = sub-stellar angle = 0), we are in full symmetry, and only the diagonal elements of the matrix have a weight. So instead of running 100 (10 x 10) simulations, simulations only need to be run for the diagonal elements (10). This is shown in middle image on the right side of Figure~\ref{fig:samplingvisual}. The full disk can be divided in 10 concentric rings, since each of these regions share a common incidence / observer angle.  In the more general cases, a particular subsample choice leads to a set of distinct incidence/observer angle regions; based on the number of angles sub-divisions, PSG then chooses how many simulations are needed and the weight for each sub-region. For example, since these angles range from 0 to 90, a subsample choice of N=5 would lead to a sub-division with bins of 18° and 22 distinct radiative-transfer regions, as shown in the right most image in Figure~\ref{fig:samplingvisual}.  
 
 Using this technique, PSG is able to accurately subsample planets' heterogeneous geometries in single simulations - but with an important implicit assumption of homogeneous atmospheric/surface properties in each subsample region.  The ability to leverage the symmetry of observational properties in this averaging scheme is based upon the assumption that the radiative properties of a region vary smoothly across the region - versus potential spatial variations due to inhomogenous clouds or spikes in surface albedo.  While the number of subsamples chosen using this method can be increased in order to capture the potential effects of inhomogeneities, this comes at a higher computational expense. Instead, inhomogenous atmospheric/surface properties can also be simulated using GlobES (Global Exoplanet Spectra, \url{https://psg.gsfc.nasa.gov/apps/globes.php}) or via the API to call simulations for different regions and then weighting them appropriately.
 
 In order to validate and compare the performance of the subsample method to the Chebyshev--Gauss method, we performed radiative transfer simulations for a Jupiter-like atmosphere employing both methods at different sampling resolutions. We used profiles for atmospheric temperature, pressure, vertical mixing ratio and cloud density from \textcite{2018ApJ...858...69M} (subsequently referred to as M18); we also used the same opacity catalog as M18 to ensure that no deviations resulted from different opacity assumptions. We produced simulated spectra and examined comparisons for both a cloudy and cloud-free Jupiter simulation, in order to examine how the differences between the two methods propagated into realistic simulation. Additional details of these validation simulations are given in Section~\ref{subsec:validation}. 
 
 \begin{figure*}[b]
  \centering
  \includegraphics[scale=0.62]{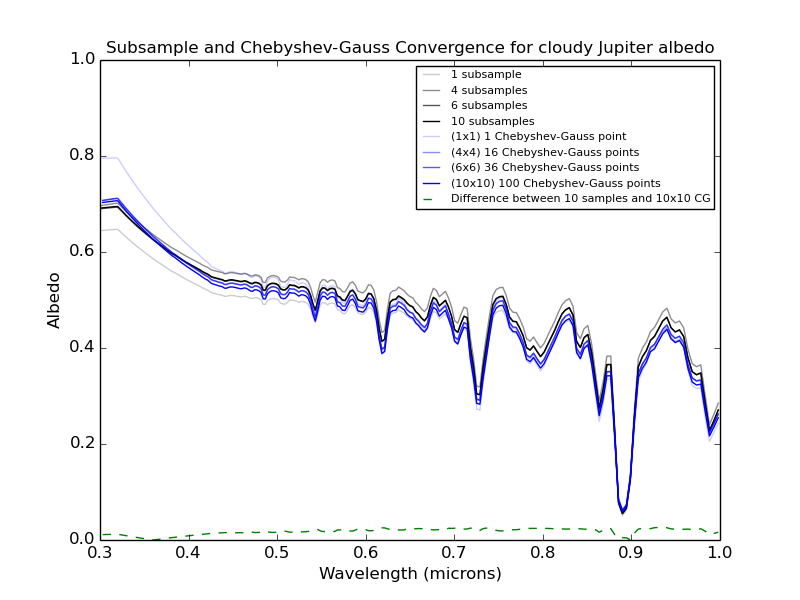}
  \caption{Simulated haze-less spectra for Jupiter with water and ammonia clouds using a range of nodes and samples for the Chebyshev--Gauss and disk subsample methods.  The two sampling methods produce very similar spectra and converge to a similar solution with an increased number of nodes/samples, as is evident from the line showing the absolute value difference between the 10 subsample and 10x10 Chebyshev Gauss points cases.  Atmospheric temperature/pressure and composition profiles are taken from \textcite{1981JGR....86.8721L} and \textcite{2018ApJ...858...69M}.}
  \label{fig:samplingjupcloudy}
\end{figure*}
 
 Results of sampling resolution tests for the Jupiter simulation with clouds are shown in Figure~\ref{fig:samplingjupcloudy}, and results for the cloud-free simulations were also very similar.  As is evident from the figure, albedo calculations for both the Chebsyshev--Gauss and subsample method converge to very similar values with a relatively small number of nodes/samples. The accuracy of the simulations using the PSG subsample method is within 5\%, 3\% and 0.2\% of the integrated flux respectively for the 1 sample, 4 sample and 6 sample simulations relative to the 10 sample simulation.  The accuracy of the simulations using the Chebyshev-Gauss method is within 5\%, 1.5\% and 1.5\% of the integrated flux respectively for the 1$\times$1 (1 node), 4$\times$4 (16 node) and 6$\times$6 (36 node) simulations relative to the 10$\times$10 (100 node) simulation. The accuracy of particular Chebyshev--Gauss and subsample node selection relative to each other is within 3\% for the integrated flux for the 10$\times$10 (100 node) and 10 subsample case. These differences in the average and integrated flux between the two methods are small -- and are likely due to differences in which they sample the atmosphere for radiative transfer calculations.  For example, in both the subsample and Chebyshev-Gauss methods implemented in PSG, there is a finite region which is used to calculated flux for each region (the subsample region and the annular beam of the Chebyshev-Gauss point) before summing or weighting. This discrete region will produce small variations in the flux that become less significant with a larger number of samples (particularly for the Chebyshev-Gauss case as sampling increases towards the limb).  In addition to that, the subsample method's averaging algorithm will inherently converge to a limit with a large number of samples, which is also evident from the comparison in figure \ref{fig:samplingjupcloudy}.

 Finally, the subsample method was also used in the validation case shown in Figure~\ref{fig:karkoschka}, where PSG was used to simulate a disk integrated Jupiter albedo spectrum from \textcite{1994Icar..111..174K}.  As discussed in the validation section (section \ref{subsec:validation}), the simulation was able to reproduce the Jupiter spectra to a great degree (with albedo variations of $\sim$0.026 on average, see the difference as a function of wavelength in the right panel of figure \ref{fig:karkoschka}), with small differences likely driven by inhomogeneous variations in atmospheric profile and cloud and haze properties as a function of location on Jupiter in both latitude and longitude.  Based on these tests, the subsample method appears to be able to accurately simulate reflectance spectra, similar to the Chebyshev--Gauss method.

Computational time for the subsample method is less expensive than the time required for Chebyshev--Gauss sampling in PSG (see Table~\ref{table:compexpense}).  While this is to be expected given that the technique was developed to take advantage of symmetries in the emergent flux from a planet, we also ran comparisons of the computational expense of the two methods for several different cases.  We ran these simulations on a personal laptop workstation with the following computer specifications: Macbook Pro with 3.1 GHz Dual-Core Intel Core i7 processor.  Computer code calling the API to PSG for both methods was written in Python. We calculated the time comparisons for the previous simulations (a cloud-free Jupiter case, a cloudy Jupiter case and a case where a haze was added to the cloudy Jupiter profile) using different subsamples. In all three cases the relative computational expense was generally the same.  The cloudy Jupiter simulations are given in Figure~\ref{fig:samplingjupcloudy} and the corresponding computational expense of the runs (as well as the other cases) are given in Table~\ref{table:compexpense}. The lower computation time required for the subsample method approaches approximately an order of magnitude as greater number of samples are used, where  computational expense of the subsample method is $O(n log(n))$ while Chebyshev--Gauss is $O(n^{2})$.  Given the similar accuracy of the subsample method and significant gain in computational expense, we use the subsample method in simulations described in the rest of the paper.  
 
\begin{table}[]
\begin{tabular}{|c|c|c}
\hline
\multicolumn{3}{|c|}{\textbf{Time Required to Complete Integration}}                                               \\ \hline
\textbf{\# of nodes/samples} & \textbf{Chebyshev-Gauss (sec)} & \multicolumn{1}{c|}{\textbf{PSG subsamples (sec)}} \\ \hline
\multicolumn{3}{|c|}{\textbf{Haze}}        \\ \hline
1  & 38.76   & \multicolumn{1}{c|}{36.94}  \\ \hline
4  & 674.22  & \multicolumn{1}{c|}{79.70}  \\ \hline
6  & 1442.96 & \multicolumn{1}{c|}{110.11} \\ \hline
10 & 4140.10 & \multicolumn{1}{c|}{167.74} \\ \hline
\multicolumn{3}{|c|}{\textbf{Cloudy}}      \\ \hline
1  & 37.05   & \multicolumn{1}{c|}{37.15}  \\ \hline
4  & 596.90  & \multicolumn{1}{c|}{85.21}  \\ \hline
6  & 1339.91 & \multicolumn{1}{c|}{113.64} \\ \hline
10 & 4284.44 & \multicolumn{1}{c|}{202.27} \\ \hline
\multicolumn{3}{|c|}{\textbf{Cloud-free}}  \\ \hline
1  & 23.92   & \multicolumn{1}{c|}{24.19}  \\ \hline
4  & 391.48  & \multicolumn{1}{c|}{28.10}  \\ \hline
6  & 889.86  & \multicolumn{1}{c|}{29.65}  \\ \hline
10 & 3086.58 & \multicolumn{1}{c|}{34.33}  \\ \hline
\end{tabular}
  \caption{Computational Expense of PSG's subsampling method versus Chebyshev-Gauss Sampling using identical input files on the same machine. Differences between the cloud-free case and the cloudy/hazy case is due to the higher requirement in scattering Legendre polynomials (LMAX) and number of stream pairs (NMAX).  Additional information on these tests and the relative accuracy of the methods is given in section \ref{subsec:subsample}.}%
  \label{table:compexpense}
\end{table}
 

\section{Validating PSG Simulated Reflection Spectra: The Jupiter Test Case} \label{subsec:validation}

\begin{figure*}
  \centering
  {\includegraphics[scale=0.55]{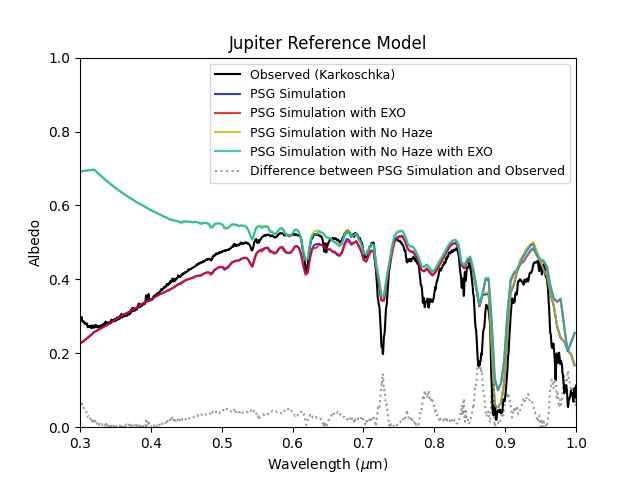}}
  \includegraphics[scale=0.55]{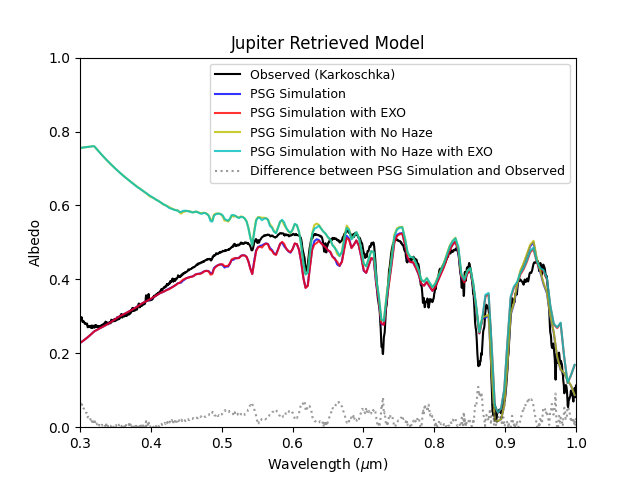}
  \caption{Simulated Jupiter spectra including haze using PSG that is matched to observations \citep{1994Icar..111..174K}.  The left image  plots simulations of Jupiter's albedo using an atmosphere that combines profiles from \textcite{1981JGR....86.8721L, 2018ApJ...858...69M}, but matches the methane abundance to that reported in \citep{2004jpsm.book...59T}.  The haze is a tholin haze with haze properties guided by \textcite{2004jpsm.book...79W}.  Simulations were carried out both with the haze and without, and were run using two different opacity database sources (HITRAN \citep{2017JQSRT.203....3G} database for the PSG simulations and the EXO database \citep{2008ApJS..174..504F, 2014ApJS..214...25F, 2017PASP..129d4402K}). The right image is a series of simulations with similar parameters as the left except with an increase in the methane abundance by a factor of 2 and increases the ammonia cloud by a factor of 2 (less constrained) with respect to M18. Both cases are able to generally reproduce the observed Jupiter spectra, especially with respect to the broad Rayleigh and haze induced continuum as well as most absorption signatures.  However, the fits do vary with respect to molecular absorption, particularly with respect to methane absorption signatures in redder parts of the spectrum}.  In both cases, the simplified models are unlikely to capture all of the complexity of the Jupiter atmosphere that is responsible for the observed albedo. 
  \label{fig:karkoschka}
\end{figure*}

In order to validate PSG's reflected light spectra simulator, we attempt to match a spectrum obtained for Jupiter's disk averaged albedo using a realistic atmosphere profile and PSG's radiative transfer suite.  The config file for this simulation is included in the appendix.  The validation spectra we use is ground-based visible spectra of Jupiter's disk averaged albedo taken using the European Southern Observatory \citep{1994Icar..111..174K} (which we refer to as `Karkoschka'). The spectra is plotted in black in Figure~\ref{fig:karkoschka} and ranges from 300$-$1000 nm, with a spectral resolution of 1 nm. The spectra is characterized by a broad continuum due to the presence of Rayleigh scattering and water and methane clouds, that is significantly darkened in the blue parts of the spectrum by haze. This broad continuum is punctuated by water and methane absorption that becomes more prominent in the redder parts of the spectrum. 
For our spectra simulations given in Figure~\ref{fig:karkoschka}, we use a haze-free model atmosphere profile from M18, which uses a derived temperature profile obtained by the Voyager spacecraft \citep{1981JGR....86.8721L}.   Pressure, metallicity (3x solar) and the consequent profile of gas abundance and water and methane clouds are all taken from M18, which is in turn based on a self-consistent radiative-convective equilibrium, chemical equilibrium model and a cloud model based on \textcite{2001ApJ...556..872A}. 
Surface properties are also largely taken from observationally driven models \citep{1981JGR....86.8721L, doi:10.1029/2005JE002411} and are given in the config file in the appendix.

Aside from the use of the PSG radiative transfer tools, our simulations of the Jupiter spectra differ from M18 primarily through the use of the sub-sampling technique described in Section~\ref{subsec:subsample} versus Chebyshev-Gauss integration and our attempts to match the spectra in \citep{1994Icar..111..174K} with the inclusion of a haze.  We also run models which examine radiative transfer using different molecular/atomic databases (with their own corresponding linelists to be used for these molecules) that influence the opacity calculations.   We specifically use databases that largely rely on the HITRAN database \citep{2017JQSRT.203....3G} for the PSG simulations, but also run simulations that largely use the EXO database \citep{2008ApJS..174..504F, 2014ApJS..214...25F, 2017PASP..129d4402K} due to both databases' widespread use in exoplanet and planetary science studies. 

The selection of parameters for the modeled stratospheric haze was based upon realistic candidates and properties in literature \citep{2004jpsm.book...79W} that was then adjusted to match the general morphology of the observed Karkoschka spectra.  We chose a Tholin haze with an effective particle radius of 0.011 \bm{$\mu$}ms and an abundance of approximately 1 part per billion from about 50 to 80 millibars (values are also available in the config file in the appendix). Composition was chosen from some of the listed chromophore candidates given in Table 5.3 of \textcite{2004jpsm.book...79W} based upon how well they fit the observed spectra. Particle size and abundance properties were also generally based off the same reference and studies discussed therein; however, differences from more sophisticated modeling that treat haze particles as aggregates of small monomers are inevitable given the simplistic treatment of the haze in these simulations.  Indeed, the significant body of literature that analyzes observations and that models the Jupiter atmosphere suggests that hazes likely vary in composition, particle size, abundance, and layer thickness both vertically and horizontally.  Additionally, haze properties also appear to considerably vary from equatorial to polar regions on Jupiter. The considerable uncertainty that remains regarding haze composition and properties suggests that high fidelity models that attempt to match spectra likely require multiple hazes with different properties that are included as a function of longitude and latitude on Jupiter.  This is beyond the scope of this paper and is also likely too detail-rich a modeling effort with respect to actual near term exoplanet observations, which will inherently be unable to resolve many of the complicated atmospheric properties the planets they observe may possess \citep{2016AJ....152..217L}.  As a result, we use a relatively simple model to match the Jupiter spectra in Figure~\ref{fig:karkoschka}, and generally note that even those measurements beyond the immediately near-term observations of reflected light from directly imaged exoplanets are unlikely to provide the signal and resolution required to resolve many of the details of a Jupiter-like atmosphere.    

Given the selected haze profile and the additional modeling choices, we are able to obtain close matches to the Jupiter spectra observed in Karkoschka, which compare favorably with past efforts that also use Jupiter spectra for validation.  Obtaining a spectra that matches both the general morphology of the data as well as specific absorption depths required variation of abundances of some of the gas molecules and cloud particle abundances in the atmospheric profile (from 0.2\% to 0.4\% for methane gas abundance and from 0.1\% to 0.2\% for ammonia cloud abundances). Two different examples of simulated spectra and their fit relative to the data are given in Figure~\ref{fig:karkoschka}. In the left panel of Figure~\ref{fig:karkoschka}, the atmosphere profile was adjusted (from a methane abundance of 0.154\% to the 0.2\%, which strengthened methane absorption features) to match the observed methane abundance for Jupiter \citep{2004jpsm.book...59T}, and the original M18 ammonia cloud profile. The simulation produces a fairly good match to the Jupiter spectra with larger deviation on some of the molecular signatures. The right panel uses an increased ammonia cloud abundance (which is less well constrained) and a volume mixing ratio for methane twice the amount reported in \citep{2004jpsm.book...59T}, producing a better match to the data.  The average difference between the model and the reference Jupiter spectra decreases from 0.37 to 0.26, and maximum difference decreases from 0.18 to 0.11 (in a methane absorption region) from the left panel to the right. The difference across the entire wavelength range is plotted in both panels and the smaller differences across the entire spectra for the right panel underlies these statistics. The simulation config file is included in appendix. 

These differences in the matching between different methane abundances are not surprising, in particular due to the lack of accurate opacity data for methane at these wavelengths for a broad range of temperatures and pressures, and also considering the relatively simple aerosols model assumed here. As alluded to before and in other works, this should likely serve as a caution with respect to over-interpretation of exoplanet spectra when the data is relatively low-resolution, low-SNR and insufficiently sampled in time compared with appropriate radiative and advective timescales.  Spatial heterogeneity of aerosol properties due to convection, variations in condensate formation, supply, transport and coagulation and dynamics-driven heterogeneities in temperature and abundances are likely to result in complexity in an atmosphere that is difficult to capture in our own solar sytem, and which will likely be more difficult for exoplanets \citep{2014arXiv1403.4436F}.  The long integration times (on the order of 10-100's of hours - for example, see the 400 hour spectroscopy limit for Roman from \citep{2019arXiv190104050B}) currently required, even compared to just the rotation periods of Jupiter and Saturn, suggest that signatures of spatial variability on directly imaged planets will be averaged out and lost for exoplanets observed in the near future.

As noted in previous sections, we also tested agreement between PSG and other simulated spectra, with a specific focus on the Jupiter spectra simulated in M18.  Figure~\ref{fig:samplingjupcloudy} shows a representative comparison of those tests, where there is strong agreement between the spectra simulated using the subsample method in PSG with simulations from that work \citep{2018ApJ...858...69M}.  Finally, the current template Jupiter atmosphere profile in PSG, taken from  \citep{doi:10.1029/2005JE002411}, was also simulated and compared to the observational data.  Again, there was broad agreement between the two, with differences likely a result of the relative simplicity of the simulated model. 

\begin{figure*}[t!]
  \centering
  {\includegraphics[scale=0.45]{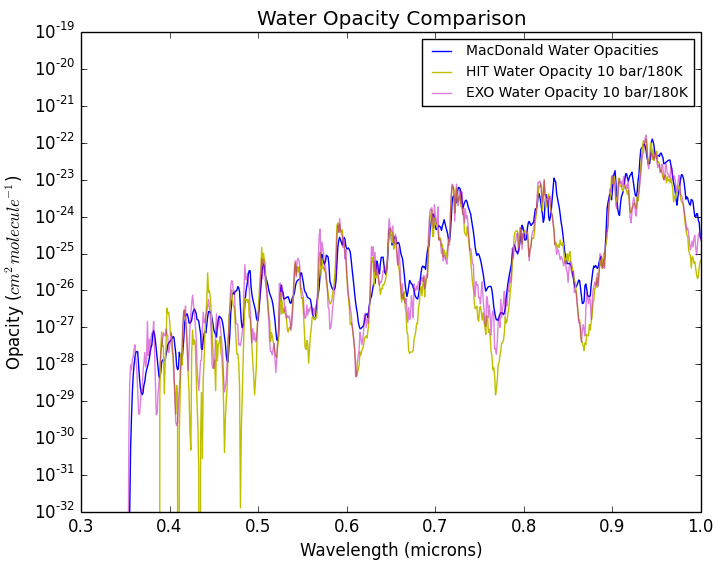}}
  \includegraphics[scale=0.57]{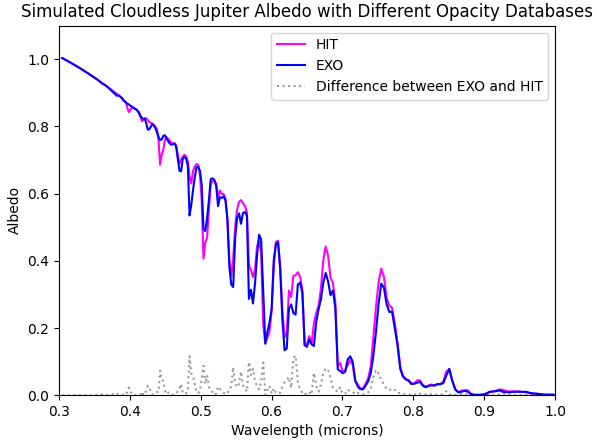}
  \caption{The left panel is a plot of the water opacity and the noticeable differences between three different sources: The HITRAN database, EXO database and the opacities used by M18.  The right panel shows the albedo spectra of a simulated cloudless Jupiter atmosphere for two different cases.  One uses the HITRAN database for key molecules while the other uses the EXO database.  While the spectra generally agree, there are noticeable differences that correspond to wavelengths that exhibit differences in opacity for the databases.   }
  \label{fig:opacity}
\end{figure*}

\section{The Importance of Opacity Table Selection on Reflection Spectra Simulations} \label{sec:opacitytables}

We also briefly comment on the effect of using different molecular/atomic databases (with their own corresponding linelists to be used for these molecules) in the simulation of spectra.   We specifically use databases that largely rely on the HITRAN \citep{2017JQSRT.203....3G} database for the PSG simulations and also run simulations that largely use the EXO database \citep{2008ApJS..174..504F, 2014ApJS..214...25F, 2017PASP..129d4402K} in most of the work in this paper.  However, each database has differences between molecules that can then propagate into the simulated spectra. Methane is particularly problematic, since there is no accurate and complete linelist that samples this range, and the EXO and PSG databases rely on decades-old UV/visible methane cross-sections taken for a single pressure and temperature. As a further example of the differences that exist for certain key molecules, in the left panel of Figure~\ref{fig:opacity}, we plot opacity for water for both the HIT and EXO databases as well as the opacity used for water in M18 (taken from Figure 1 of M18).  Differences between the three are fairly clear in that image, but are not necessarily obvious with respect to their effect on spectra.  In the right panel of the same image, simulations carried out in PSG for a cloudless Jupiter atmosphere are overlaid for two runs, one using HIT for key molecules and the other using EXO, as is the difference between the two simulations.  In figure \ref{fig:opacity}, the general morphology of the spectrum is the same and for the most part absorption by molecules is similar (the average difference between the two cases 0.015 and the maximum difference is 0.12). However, water-related absorption is noticeably different in certain regions, with the difference approaching 10--20\% of the albedo at particular wavelengths. 

The effect is more muted when clouds and hazes are included as they typically tend to reduce the influence of absorption due to molecules present in the atmosphere. Small differences are evident in Figure~\ref{fig:karkoschka} between similar cases that are run using PSG (which heavily relies on HIT) versus those that rely heavily on EXO, but in these simulations of a hazy atmosphere of Jupiter, it is clear that the differences are far smaller than the cloud-free case and likely largely irrelevant to exoplanet observations. However, the differences between databases should not be ignored, and given the potential to impact simulations and interpretation of the albedo spectra of relatively clear atmospheres, should be a topic of greater scrutiny. A quantified comparison of the difference between different opacity tables may be useful in the future in order to extricate the potential effects on interpretation of observations. 

\section{Simulating a Promising Nancy Grace Roman Space Telescope Target: Simulated Spectra of $\upsilon$ Andromedae d} \label{sec:UpsAndd}

In addition to examining general questions regarding issues related to forward models of planetary reflectance spectra, we also examined the potential science questions that could be investigated for the bright Roman Space Telescope fiducial target, Upsilon Andromedae d (subsequently referred to as $\upsilon$ And d). $\upsilon$ And d is a $10.25^{+0.7}_{-3.3}$ $M_{Jup}$ planet \citep{2010ApJ...715.1203M} with a derived semi-major axis of 2.53 AU that orbits a 1.27 solar-mass F8V star \citep{1998A&A...336..942F}.  Joint fits using radial velocity and astrometry data in modeling \citep{2010ApJ...715.1203M} of planets c and d have been used to constrain orbital parameters and indicate that $\upsilon$ And d is on an inclined and eccentric orbit with i=23.758 and e=0.316. There is also evidence for a 4th planet in the system (see \textcite{2010ApJ...715.1203M, refId0}), but it is important to note that orbital and planetary parameters for the potential planet are unconstrained given complications arising from observational systematics \citep{2015ApJ...798...46D}.  This is notable for our purposes as orbital and planetary parameters for all the planets in the system, including $\upsilon$ And d, are dependent on joint fits. As a result, we use the parameters from \textcite{2015ApJ...798...46D} for the purposes of simulating potential spectra. 

\subsection{$\upsilon$ Andromedae d as a Nancy Grace Roman Space Telescope Target} \label{subsec:UpsAnddNGRST}

$\upsilon$ And d is an attractive target for the Roman mission and other future efforts due to its favorable observational properties. The 2015 Roman Science Definition Team report \citep{2015arXiv150303757S} lists some of the most promising targets for the observatory from the sample of known
radial-velocity planets at that time.  In Table 2-7 of the report, observational parameters for some of the most favorable targets are listed, including separation from host star, relative contrast to the primary, and integration time required to obtain a signal-to-noise ratio of 5 at 565 nm with a 10\% bandpass using the Coronagraph Instrument (CGI) Hybrid Lyot Coronagraph (HLC).  While the integration times made simple assumptions for planetary albedo and noise contributions, the relative values give a useful measure of the potential efficacy of observations of different planets.  Due to the $\upsilon$ And A system's distance and host star's brightness, the integration time required for observations of $\upsilon$ And d to reach the required SNR is one of the shortest among all of the listed planets.  The angular separation for $\upsilon$ And d is fairly small at 0.1805 arcsec for mean orbital separation, but is larger than all of the planets with shorter integration times in table 2-7 from \textcite{2015arXiv150303757S}, which makes it an advantageous target with respect to inner working angle considerations relative to brighter planets. The contrast ratio for $\upsilon$ And d is also favorable and is better (using the assumptions listed in the report) than other common targets of simulation, such as 47 UMa c and $\upsilon$ And e.  Indeed, all planets in table 2-7 from \textcite{2015arXiv150303757S} with better contrast ratios are at shorter angular separation $<0.16$ arcsec from their host star than $\upsilon$ And d.  Some early simulations (described in section \ref{subsec:prevsims}) of SNR also reinforce the observational favorability of $\upsilon$ And d. Simulated SNR predictions using Jupiter-like spectra for all the then-known RV planets at a number of different bandpasses also demonstrated the relatively high SNR that could be obtained for $\upsilon$ And d \citep{2019AJ....157..132L}.  This is especially true at the bluer wavelengths relevant to the current imaging and spectroscopic capabilities of Roman, (though \textcite{2019AJ....157..132L} found that at redder wavelengths, the assumed spectra for $\upsilon$ And d yielded less favorable SNR).  This combination of observational qualities motivates us to simulate a range of different $\upsilon$ And d spectra that reflect different atmospheric states (given in table \ref{tab:upsanddparam}), in order to help plan for potential future observational programs. 

\subsection{Previous Spectral Simulations of $\upsilon$ Andromedae d} \label{subsec:prevsims}

There have been some earlier simulations of the spectra of $\upsilon$ And d.  \textcite{2019AJ....157..132L} was the first study to specifically examine SNR for $\upsilon$ And d assuming both a Jupiter and Neptune-like spectra, but earlier work by \textcite{Sudarsky_2000, Sudarsky_2003} did try to produce generic spectra for the class of planets that $\upsilon$ And d may fit into.  These works used atmospheric structure and radiative transfer modeling to produce broad extrasolar giant planet classifications based on qualitative similarities in composition and spectra of objects within broad effective temperature ranges.  In \textcite{Sudarsky_2003}, $\upsilon$ And d was modeled as a `Class II' `water class' planet characterized by tropospheric water clouds.  Simulations propagated the `Class II' planet spectra using the system and orbital constraints at that time.  While the spectra produced were for a Jupiter-fixed orbital radius, phase averaged planet, \textcite{Sudarsky_2003} did acknowledge that the substantial eccentricity of the planet means it may be too warm in its outer atmosphere to possess water clouds at periastron, and cross over into the water-cloud-free `Class III' planet regime.  This potential variation in planetary and atmospheric parameters due to $\upsilon$ And d's eccentric orbit is just one of the interesting facets of the system that suggests more detailed modeling may be warranted.  

Planets such as 47 Uma c and $\upsilon$ And e may be more similar to and are often modeled as Jupiter analogues, but $\upsilon$ And d represents a giant exoplanet that exists in an atmospheric phase space not seen in the solar system (for example, see Figure 1 of \textcite{2016AJ....152..217L}).  Its orbit places it in the conventional liquid water habitable zone of its system \citep{2006Icar..183..491B}.  The variation in separation due to eccentricity may mean that the planet experiences significant variation in the existence and location/morphology of clouds in its atmosphere, which may modify its albedo.  The planet may transition from a relatively cloud-free `Class III' giant planet at periastron to a water-cloud-possessing `Class II' planet at apastron, which would result in a shift from a relatively low-albedo average reflectance to a high albedo average reflectance (indeed, \textcite{2000ApJ...538..885S} suggests planets with high water clouds may be brighter in the visible than Jupiter analogues); brightening would occur as the planet's orbital separation approaches a maximum, which would also make it less likely to be interior to the inner working angle of the Roman CGI coronagraphs. The variation in temperature and consequently cloud location as a function of phase may therefore produce a measurable effect on the planetary spectra.  Finally, as is the case of gas giants in the solar system, $\upsilon$ And d may possess hazes that alter its spectra.  Given the location of $\upsilon$ And d and the common presence of moons and subsequently the in-fall of their material on their giant planet hosts in the solar system, such hazes that may contain albedo-influencing chromophores that may be either exogenously or endogenously sourced.  This suggests that there are a number of different types of potential atmospheres that $\upsilon$ And d may possess, which guides our simulations of the planets' spectra in the rest of this section. 

\subsection{Determining Key Parameters for Simulations of $\upsilon$ Andromedae d} \label{subsec:keyparams}

\begin{figure}
\CenterFloatBoxes
\begin{floatrow}
\ffigbox
  {
  \includegraphics[scale=0.48]{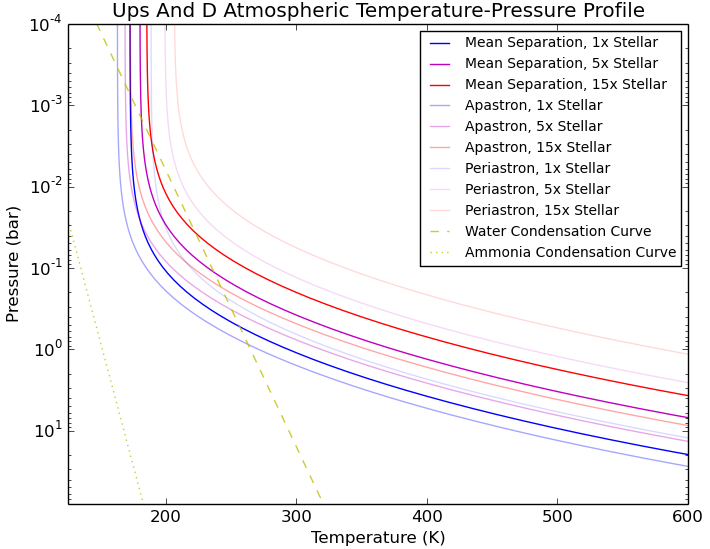}}
  {\caption{Atmosphere temperature/pressure profiles for $\upsilon$ And d using parameters (for 1/5/15x stellar metallicity \citep{2007MNRAS.378.1141G}) that are a subset of those tested for the planet at periastron, apastron and mean orbital separation.  The overlaid dashed and dotted lines are water and ammonia condensation curves, respectively, for Jupiter-like gas abundances.  Intersections with the planets atmospheric profile lines indicate potential locations for clouds.  }\label{fig:upsanddtp}}
\killfloatstyle
\ttabbox
  {\begin{tabular}{cc} 
  \hline \hline
  $\upsilon$ And d Parameters \\ \hline \hline
  $^{1}$Observation Derived \\ \hline
  Planetary Mass - $10.25^{+0.7}_{-3.3}$ $M_{Jup}$ \\
  Planetary Radius - $1.02$ $R_{Jup}$ \\
  Semi-major Axis - 2.53 AU \\ 
  Orbital Eccentricity - 0.316 \\
  Orbital Inclination - 23.758\degree \\
  \hline
  Model Parameters \\ \hline
  $T_{eff}$ (max, min, mean), $T_{int}$=0  -  260K, 188K, 215K \\
  $T_{eff, mean}$ inc. $^{2}$$T_{int}$ ($10.25  M_{j}$, $6.95  M_{j}$) -  319K, 270K \\
  Planetary Gravity (\textit{g}) - 244.23 m/s$^{2}$  \\ $^{3}$Metallicity ([Fe/H]$_{star}$ = 0.131) - 1/5/10/15x \\
  Water Cloud Particle Size - 0.1/1$\mu$m \\
  Haze Properties - no haze/Tholin Haze \\
  \hline  
  $^{1}$ \citep{2010ApJ...715.1203M, 2015ApJ...798...46D} \\
  $^{2}$ \citep{marley_mark_2018_1309035} \\
  $^{3}$ \citep{2007MNRAS.378.1141G}
  \end{tabular}
  }
  {\caption{$\upsilon$ And d parameters used for spectra simulations.}\label{tab:upsanddparam}}
\end{floatrow}
\end{figure}

In order to simulate spectra of potential atmospheres of $\upsilon$ And d, we first obtain an atmospheric temperature and pressure structure for the planet.  This is based on effective temperature and planetary gravity parameters that are dependent on planetary radius.  Since there are no observational constraints for the radius, we use models that explore the mass--radius relation for giant planets in order to produce an estimate for the radius.  While the radius is a function of a number of different variables, including age of the system and core size \citep{2007ApJ...659.1661F, 2008ApJ...683.1104F}, most models of the mass--radius relation find that the radius of evolved gas giants in the range of Jupiter's mass are not likely to exhibit large radius variations with mass \citep{2017A&A...604A..83B}.  Using relations \citep{2007ApJ...659.1661F, 2008ApJ...683.1104F} for evolved planets with masses and orbital separations relevant to $\upsilon$ And d, we use a value of 1.02 $R_{Jupiter}$ for the planet.  This yields a value of g$\sim$245 $\text{m/s}^{2}$, which is an order of magnitude larger than Jupiter.  With these values we can the use the analytical fits provided by M18 in order to produce a temperature and pressure profile for ups And d.  M18 created these by fitting a number of self consistent T/P models using the methods described in \textcite{2008ApJ...683.1104F}.  While the models use a maximum of 100 $\text{m/s}^{2}$ for the planetary gravity, we test lower values of gravity between that value and the calculated value for $\upsilon$ And d and find it does not result in a significant difference in the profile and potential locations of cloud formation. Due to the substantial eccentricity and subsequent variation in orbital separation from its host star that $\upsilon$ And d possesses, it is possible that the atmospheric profile for the planet may vary from periastron to apastron.  A calculation of effective temperature ($T_{eff}$) indicates a variation of $\sim$70 K from periastron to apastron with a mean $T_{eff}$ of 215 K.  In this case, we do not include the potential contribution of internal heating ($T_{int}$) to $T_{eff}$ given uncertainty of $\upsilon$ And d's internal properties. The value of $T_{eff}$ for the solar system giant planets is approximately 10-20 K greater than their $T_{eq}$ - such an increase would not change the $T_{eff}$ of $\upsilon$ And d into a significantly different condensation regime for apastron and mean separation. For periastron, such an increase would potentially result in a cloud-free atmosphere at larger metallicity values.  In order to explore the potential effect of this variation on the atmosphere, we plotted a T/P profile for the planet (with several metallicity values for the planet as a function of the stellar metallicity for $\upsilon$ And A) for the corresponding effective temperatures at periastron, apastron and its mean orbital separation (listed in Table~\ref{tab:upsanddparam}).  The profiles are given in Figure~\ref{fig:upsanddtp}, with condensation curves for water and ammonia overlaid.  While the profiles at the orbital extrema are unlikely to exactly match the actual profile given the latency inherent in the structure of the atmosphere due to past conditions,  these profiles give general guideposts for the atmospheric parameters that may control observables.  We do also include simulations (given in Table \ref{tab:upsanddparam}) with modeled $T_{int}$ values for the constrained mass and age of $\upsilon$ And d.  Those values are based on closest values that correspond to $\upsilon$  And d in the Sonora-Bobcat tables \citep{marley_mark_2018_1309035} and are simulated for the reference mass value but also the upper and lower bound on mass for the planet.  We then use these $T_{int}$ and $T_{eff}$ values in order to simulate spectra for these cases and examine potential observations by Roman.

We also examined the general effect of gravity and metallicity on the profiles.  As discussed above, in general there was not significant variation of where profiles intersected with condensation curves as planetary gravity was varied.  At significantly higher values (2$\times$ the value for $\upsilon$ And d's observation derived mass) of gravity than the value calculated, the periastron and apastron profiles converged towards the mean separation profile with a temperature difference of less than 10K above approximately 1 bar.  Intersection of the water condensation curve with the temperature profile did not vary significantly and the main difference was that at around the 50mb level and above, temperatures were actually higher for the apastron case versus the periastron case, though the difference was less than 5K and both were close to the mean separation value (the periaston temperature decreased with higher \textit{g}). At lower values of \textit{g}, the difference in temperatures at the orbital extrema was more pronounced, as apastron temperatures decreased and periastron increased (for example, at Jupiter-like values of \textit{g} - unrealistic given observed parameters - stratospheric temperatures increased by about 20K at periastron, and apastron temperatures decreased by a smaller amount).  At these lower values of \textit{g}, the periastron profile did not intersect with the water condensation curve, suggesting a potential lack of water clouds.  

Metallicity had a more significant effect on the profiles for observed metallicities in the solar system gas giants \citep{WONG2004153, 2009Icar..199..351F}. Figure~\ref{fig:upsanddtp} shows profiles for the three orbital distances at different metallicities, with increased metallicity denoted by increasingly transparent lines (temperature/pressure profiles for additional cases with modeled $T_{int}$ values are given in section \ref{sec:TPTint} of the appendix). There are some pronounced effects of increased metallicity - notably that at metallicities near or greater than Jupiter-like enhancement values, the periastron T/P curve may suggest a water-cloud free atmosphere.  Indeed, even the mean separation T/P profile at 15x metallicity is close to a water-cloud-free case, suggesting that more enhanced metallicity values would lead to a planet where water clouds may only be favored as the planet approaches apastron (tests of much higher metallicity values - $>$25 - shows that such a planet would still likely possess a thin layer of clouds at mean separation, but that such clouds would become thinner). The actual process of formation and destruction of these clouds is beyond the scope of this paper but would be dependent on the stellar irradiation environment and sedimentation and production timescales associated with cloud formation. These differences also result in variations of the intersection of T/P profiles with the water condensation curves by over an order of magnitude in pressure for potential deeper clouds.  The variation of these profiles and subsequently the atmospheric features with realistic metallicity values necessitates models which consider a range of different atmospheric properties. 

\subsection{Composition of the $\upsilon$ Andromedae d Atmosphere} \label{subsec:composition}

In our simulations of $\upsilon$ And d's atmosphere, we first used a thermo-chemical equilibrium chemistry code in order to compute molecular abundances of the likely dominant molecules in the atmosphere.  We used the package GGchem \citep{2018A&A...614A...1W}, an open-source computer code that can determine the chemical composition of gases in thermo-chemical equilibrium down to 100 K.  GGchem allows users to choose elements and their associated abundances, sources for equilibrium constants and temperature and pressure settings when setting up a model.  This enables chemistry calculations for atmospheres of varying metallicity with a given temperature-pressure profile for different vertical layer representations.  The software also includes an equilibrium condensation code (which we do not use in this study) in addition to the gas-phase equilibrium chemistry component. The software is especially useful for $\upsilon$ And d as it can simulate equilibrium chemistry at temperature ranges ($\sim$160-700K) relevant to the planet (shown in Figure~\ref{fig:upsanddtp}).

We ran GGchem for the T/P profiles we obtained for different metallicity values using a 120-layer atmosphere that is evenly distributed in pressure logspace from $10^{-7}$ to $10^{3}$ bars.  However, our radiative transfer calculations, and consequently our simulated spectra, are based on a maximum pressure limit of 10 bars, (which is often taken as the troposphere boundary or atmosphere surface for the solar system gas giant planets \citep{1998JGR...10322857S, 2004ESASP1278..331E}). Once the model is run using a selected list of species and their abundances, an output of the most common molecules at a particular T/P value is produced. We then screened the molecules we would use for the rest of our calculations by only choosing the compounds and elements that exceeded greater than 1 part per million at the 1 bar level for the 1x or 5x metallicity cases. This leads to an atmosphere which is composed of six dominant molecules that comprise the vast majority ($>$99.999$\%$) of the total abundance:  H$_{2}$, He, CH$_{4}$, H$_{2}$O, H$_{2}$S, and NH$_{3}$. While this threshold does capture the vast majority of the modeled atmosphere composition that is likely to influence spectra, we do note that this excludes Alkali elements and compounds such as sodium and potassium, which may exhibit spectral features for higher metallicity cases.

The metallicity cases that were modeled were 1x, 5x and 15x scenarios and were chosen to encompass a range of values which reflect the current understanding of values for Jupiter and Saturn \citep{2016arXiv160604510A}.  Increased granularity of metallicity values, a wider range of values and the inclusion of more realistic metal abundances that are less simplistically correlated would all be valuable improvements for future study.  These metallicity values and the appropriate T/P profiles for each orbital position were then used to determine the existence and morphology of clouds.  As discussed above, the latency in atmospheric properties between different orbital positions may mean some of the input parameters do not instantaneously effect atmospheric structure.  However, as physically motivated fiducial models, such external parameter driven atmospheric structure can help inform how spectra may potentially change over a sequence of observations.  

\subsection{Clouds and Hazes in the $\upsilon$ Andromedae d Atmosphere} \label{subsec:cloudshazes}

 \begin{figure*}[t]
  \centering
  \includegraphics[scale=0.62]{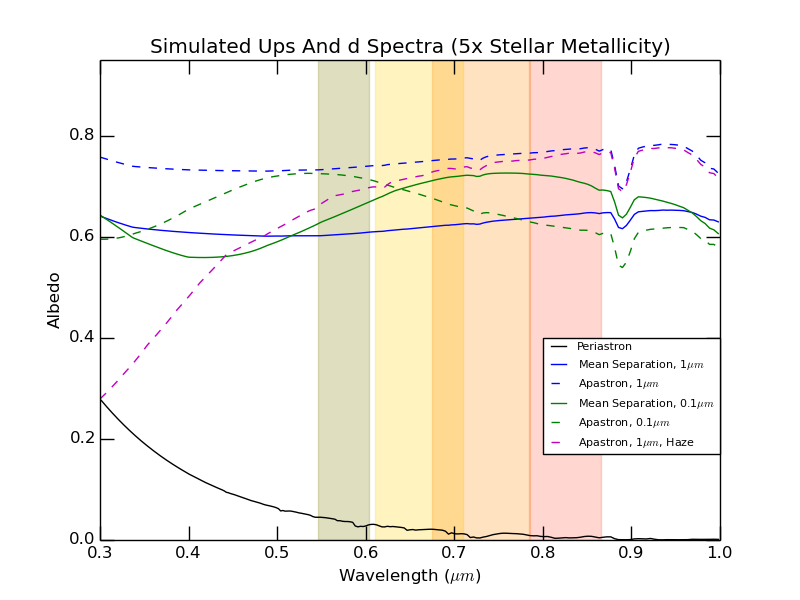}
  \caption{Simulated geometric albedo spectra at phase=0 for $\upsilon$ And d assuming a 5x stellar metallicity atmosphere.  Spectra are plotted for the planet when it is located at different orbital distances - periastron, mean separation and apastron.  The periastron case is cloud-free, while both of the other positions result in atmospheres that possess high, water clouds.  For the cloudy cases, spectra are shown for cloud particles sizes of 0.1 and 1\;$\mu$m, and a case at apastron with a tholin haze is also included}
  \label{fig:UpsAndd5xmetalcases}
\end{figure*}

Cloud thickness, vertical location and other properties were all chosen based upon physically motivated assumptions taken from modeled profiles.  Cloud extent was chosen based upon intersections of T/P profiles with condensation curves, as is visible in Figure~\ref{fig:upsanddtp}.  Total cloud mass was then chosen in order to obey mass conservation based upon the integrated condensible column above the deepest saturation layer using the equilibrium chemistry calculations of abundance of the volatile of interest. Given the relevant T/P range, this entails integrating water vapor abundance in condensation regions and then converting to water ice, with a very small, decreasing amount of water vapor retained in the profile.  Cloud thickness structure was set with a layer of maximum cloud condensate abundance ratio located where supersaturation was the highest.  The condensate mixing ratio in adjacent layers was then set by prescribing gradual attenuation in layers below the maximum layer and a steeper attenuation in layers above (with endpoints in the condensation range).  Variations in the attenuation were tested and do have some effects on spectra, but are generally at levels that do not influence the broader interpretation relevant to near term observations (for example the narrowest versus broadest maximum layer schemes had differences less than $\sim$several $\%$ but one intermediate scheme had larger $\sim5-10\%$ differences in a small portion of the spectra).  Finally, given the exploratory nature of these simulations, two particles sizes were tested for the cloud particles - 0.1 and 1\;$\mu$m. They were chosen as initial exploratory values based upon the upper atmosphere particle sizes measured by the Galileo Probe Nephelometer and to test smaller particle sizes motivated by evidence of refractory clouds on hot Jupiters \citep{2004jpsm.book...79W, 2016A&A...594A..48L}. Additional particle sizes and more realistic distributions of sizes will be important variables to test in future work. The water ice cloud properties and scattering model details are given in the config files included in the appendix, and PSG allows the user to choose from a number of options for the aerosol properties.  In the simulations described in this paper, the refractive indices for the water ice clouds were taken from \textcite{2013JQSRT.130..373M, 2017JQSRT.203....3G} while the Mie scattering implementation used was from \textcite{1983asls.book.....B} and used 20 angles and 200 size bins from 0.005 to 20 $\mu$m. The hypothetical effect of a haze layer was the last additional component included in the simulations, with a thin, high layer of tholin haze composed of 0.1 and 1 $\mu$m particles (particle size based on the approximate peak of a number of modeled haze particle distributions - see Figure 8 of \textcite{2021arXiv210203480G} (specific details on haze particles used are available in the attached config files in the appendix).

\subsection{Simulation Results: Geometric Albedo of $\upsilon$ Andromedae d} \label{subsec:geoalbedo}

\begin{figure*}[t]
  \centering
  \includegraphics[scale=0.65]{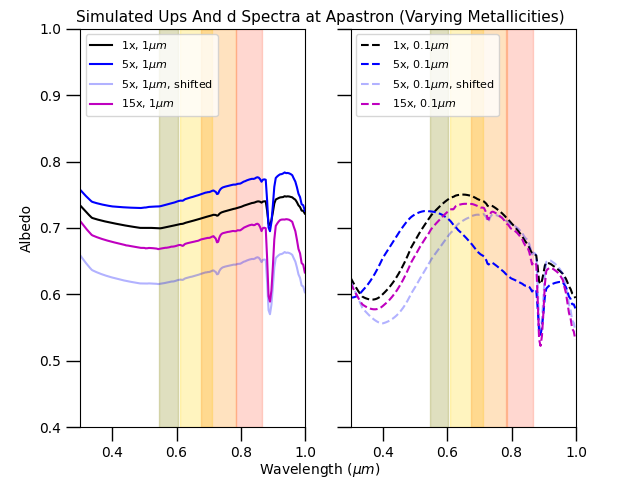}
  \caption{Simulated geometric albedo at phase=0 for $\upsilon$ And d at apastron for different metallicity cases.  The left panel shows 1x, 5x and 15x metallicity cases for 1-$\mu$m cloud particle sizes, while the right panel shows the same cases for 0.1-$\mu$m cloud particle sizes.  The `shifted' lines are plots of the albedo spectra for the 5x metallicity case if the maximum cloud fraction location was artificially shifted in order to match the flux peak in the 0.1-$\mu$m cloud particle size case.}
  \label{fig:UpsAnddmetalcompare}
\end{figure*}

Modeled geometric albedo spectra from a number of selected simulations for the $T_{int}$=0 case are given in Figures \ref{fig:UpsAndd5xmetalcases} and \ref{fig:UpsAnddmetalcompare}. Current Roman CGI filters are overlaid in olive green, yellow, orange and red.  Figure~\ref{fig:UpsAndd5xmetalcases} shows the spectra for models that all simulated a 5x stellar metallicity atmosphere for $\upsilon$ And d.  These spectra were produced for the planet at three separate orbital positions - periastron, mean separation distance from the the host star, and apastron.  These different orbital distances corresponded to different effective temperatures and consequently atmospheric profiles.  For the periastron case, the atmosphere was too warm for water clouds to condense in the portion of the atmosphere that was simulated.  The result is a cloud-free atmosphere that produces a relatively low albedo across the entire visible spectral range, with slightly higher reflectivity in the bluer wavelengths due to Rayleigh scattering. Both the mean separation and apastron cases produced atmospheres with water cloud condensation - in both cases with high water clouds that significantly increased albedo through the entire visible range. For these cloudy atmospheres, we tested both of the cloud particle sizes described above and for the apastron case, also included a model with the tholin haze.  Apastron cases are given with the dashed lines, while mean separation cases are denoted with the non-black solid lines.  Within those cases, particle size is distinguished by the color of the lines (green is used for 0.1\;$\mu$m and blue is used for 1\;$\mu$m). The justification for the choice of particle sizes is given in Section \ref{subsec:cloudshazes}. The haze has the expected effect of reducing albedo in the bluer wavelengths (by more than a factor of two at 0.3\;$\mu$m) and having little effect towards redder wavelengths.  Unsurprisingly, particle size also has significant effect on the spectra.  While the 1-$\mu$m particle size cases tend to have less variation in albedo in the spectral range examined, the 0.1-$\mu$m cases exhibit more significant variation from 0.3 to 1\;$\mu$m, with peak and trough albedo location varying by wavelength due to a combination of both the particle size and the vertical position and extent of the clouds.  The brightest albedo spectra across the entire wavelength range is in the case of the planet at apastron with water clouds composed of 1-$\mu$m particles.  Spectra with the same cloud particle size, except with the planet at mean separation, possess albedos approximately 10-20\% less bright due to changes in cloud height, extent and thickness. A summary table that details the cloud properties for these simulations is given in Section \ref{sec:cloudprop} of the appendix.  This effect of planet orbital distance is more complicated in the 0.1-$\mu$m case, as the albedo is higher at bluer wavelengths at apastron and at redder wavelengths for the mean separation case due to the different cloud structure.  Absorption signatures, mostly due to methane, are equally prominent in just about all cases, with the small variations due to the vertical location of the cloud decks.   Variations between the separate cases at a particular orbital distance (periastron/mean/apastron separations are 1.73/2.53/3.33 AU, respectively) may be distinguishable, but at least in these models, some of the most diagnostic wavelengths may be bluer than current Roman filters.

Figure~\ref{fig:UpsAnddmetalcompare} shows a comparison of model geometric albedo of the atmosphere assuming different metallicity values for when the planet is at apastron. The left panel shows how the spectra changes with varying metallicity while assuming cloud particle sizes of  1\;$\mu$m, while the right panel shows the same set of cases but with a cloud particle size of 0.1\;$\mu$m. While particular metallicity cases may exhibit broad patterns of brighter or darker continuum spectra across the entire wavelength range relative to each other, the relationship between metallicity and overall continuum brightness is not a linear one.  This is because the relationship of a particular metallicity for the atmosphere and its corresponding T/P profile does not translate to cloud formation with respect to cloud height/extent or density in a simple pattern.  Since these cloud properties (see Section \ref{sec:cloudprop} of the appendix and \textcite{2013JQSRT.130..373M, 2017JQSRT.203....3G} for details) all have significant and complicated effects on spectra, continuum brightness can vary in non-intuitive ways.  For example, in the 1-$\mu$m cases, the continuum brightness of the thicker cloud deck in the 1x metallicity case is greater than that of the 15x metallicity case.  However, the 5x metallicity case is the brightest with respect to continuum albedo.  The only relatively intuitive relationship is that of the strength of the absorption signatures, which increase with increasing metallicity due to higher molecular abundances.  In the 0.1-$\mu$m cloud cases, the non-intuitive cloud effects for varying metallicity cases are even more pronounced, as the peak albedo for the 5x metallicity case is offset (as is the entire spectra) from the 1x and 15x metallicity cases. This difference (and perhaps the differences in the 1-$\mu$m case) may be due to the location of where the cloud layers in each case become opaque. For example, the 1x and and 15x metallicity cases experience a maximum cloud fraction at 0.058 and 0.024 bars respectively, while the 5x metallicity case has maximum cloud fraction at 0.069 bars.  The deeper atmosphere cloud appears to account for the difference in the 0.1-$\mu$m curve peaks - artificially setting the highest fraction at 0.344 bars for the 5x metallicity case instead leads to a similar morphology to 1x and 15x cases, as is demonstrated by the `shifted' blue line in the 0.1-$\mu$m case.  The effect of this demonstration of the impact of maximum cloud fraction and its corresponding effect on opacity and the spectra is also included in the 1-$\mu$m case, where the 5x metallicity case goes from the brightest continuum albedo (approximately 0.75) to the darkest one (approximately 0.65) in this contrived scenario.  With the current Roman CGI filters \citep{10.1117/12.2562997} overlaid in these plots as well, it is apparent that depending on signal to noise described in Section \ref{subsec:geoalbedo}, it may be possible to resolve some of these properties based upon the spectra. However, given the dramatic effect that such variations in atmospheric properties and cloud morphology can have on albedo, a more detailed study of the different atmosphere scenarios that are considered plausible is highly warranted. 

\subsection{Simulation Results: Phase and Illumination Appropriate Albedo of $\upsilon$ Andromedae d} \label{subsec:geoalbedo}

 \begin{figure*}[t]
  \centering
  \includegraphics[scale=0.65]{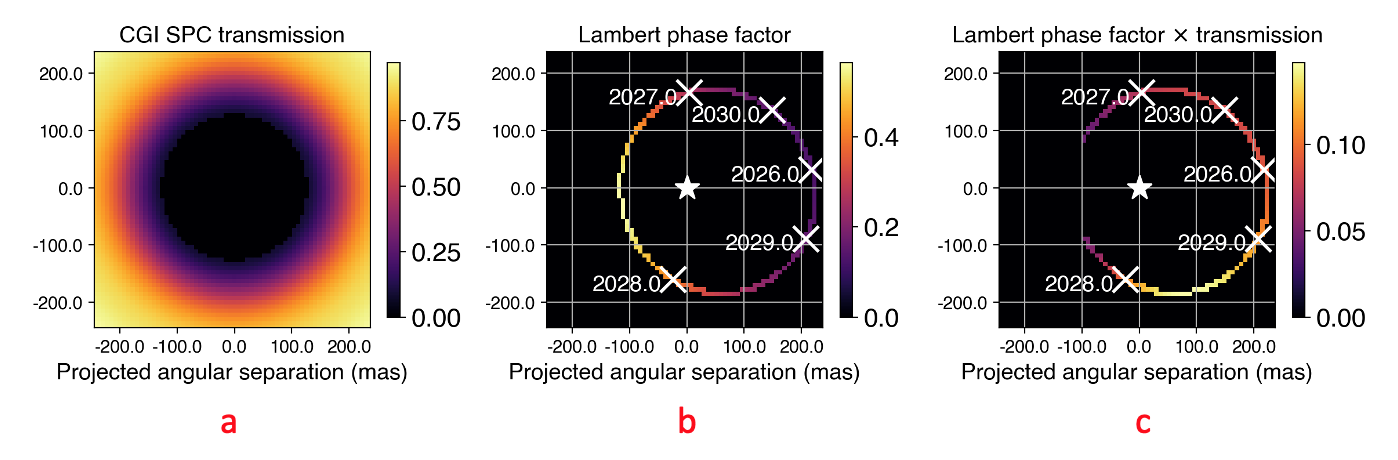}
  \caption{a) Relative PSF core transmission of the CGI Band 3 SPC coronagraph as a function of the angular offset from the occulted star. b) Predicted offset of $\upsilon$ And d, plotted for dates 2026.0--2030.0, and color-coded according to the Lambert phase factor at each position. c) The center plot repeated, but now with the Lambert phase factor scaled by the relative coronagraph transmission map, indicating the section of the orbit with most favorable observability falls in the range of dates 2028.0--2029.0.}
  \label{fig:UpsAnddEphemeris}
\end{figure*}

While the spectra in Figures \ref{fig:UpsAndd5xmetalcases} and \ref{fig:UpsAnddmetalcompare} were examples of the planet's spectra if it were viewed at full phase, a realistic simulation of $\upsilon$ And d's expected spectra requires consideration of the orbital phase and illumination state of the planet.  $\upsilon$ And d is somewhat unique in this regard, in that its 3-dimensional orbital structure has been explored due to the existence of mass constraints from RV and astrometry measurements of the system \citep{2010ApJ...715.1203M, 2015ApJ...798...46D}. We refer to spectra that incorporates existing constraints on inclination and eccentricity along with phase variation as `phase and illumination appropriate' spectra. Realistic simulations of the planet's spectrum also need to take into account the field of view-dependent throughput \citep{10.1117/12.2562997} of the instrument used to observe the system.  For the purposes of this study, we examine simulated Roman CGI observations of $\upsilon$ And d given current instrument specifications, including two of the baseline observing modes: spectroscopy in Band 3 (675--785 nm) with a Shaped Pupil Coronagraph (SPC) and broadband imaging in Band 1 (546--604 nm) with a Hybrid Lyot Coronagraph (HLC)~\citep{Mennesson2020arXiv}\footnote{\url{https://roman.ipac.caltech.edu/sims/Param_db.html}}.   

The relative point spread function transmission of the Roman CGI Band 3 SPC mode, mapped over the coronagraph field of view, is illustrated in Figure~\ref{fig:UpsAnddEphemeris}a. In Figure~\ref{fig:UpsAnddEphemeris}b, we show the sky-projected offset of $\upsilon$ And d as a function of time, color-coded by the phase function of a Lambertian sphere. Finally, in Figure~\ref{fig:UpsAnddEphemeris}c, we show the combined effect of the Lambertian phase function modulated by the field-point-dependent coronagraph throughput. Together, these indicate the most favorable dates for observing $\upsilon$ And d after the expected 2026 commissioning phase of the Roman Space Telescope span roughly from the beginning of 2028 to the beginning of 2029 (approximately 90 to 150 degrees in phase from $\upsilon$ And d's periastron). Using these orbital positions and times as a guide, we then simulate spectra of $\upsilon$ And d for a number of our atmosphere models with the appropriate phase-dependent illumination and instrument-specific throughput with PSG.

\begin{table}[b]
\begin{tabular}{|c|c|c|c|c?c|c|c|c|}
\hline
\multicolumn{5}{|c?}{\textbf{Ups And d Albedo ($\bm{0.1 \mu \text{m} \: \text{Particle Size}}$)}}                                                                                            &  \multicolumn{4}{c|}{\textbf{Ups And d Albedo ($\bm{1 \mu \text{m} \: \text{Particle Size}}$)}}                                                                                                                                       \\ \hline 
\textbf{Degrees from} & \textit{575 nm} & \textit{660 nm} & \textit{730 nm} & \textit{825 nm} & \textit{575 nm} & \textit{660 nm} & \textit{730 nm} & \textit{825 nm} \\
\textbf{Periastron} & \textit{filter} & \textit{filter} & \textit{filter} & \textit{filter} & \textit{filter} & \textit{filter} & \textit{filter} & \textit{filter}
\\ \hline
\textit{90}                                                                & 0.213                                                            & 0.206                                                            & 0.203                                                            & 0.202  & 0.233 & 0.237 & 0.241 & 0.246                                                    \\ \hline
\textit{110}                                                               & 0.170                                                            & 0.164                                                            & 0.161                                                            & 0.162     & 0.171 & 0.172 & 0.173 & 0.176                                                       \\ \hline
\textit{130}                                                               & 0.135                                                            & 0.131                                                            & 0.130                                                            & 0.133                                                        & 0.120 & 0.119 &  0.118 & 0.117   \\ \hline
\textit{180}                                                               & 0.103                                                            & 0.097                                                            & 0.096                                                            & 0.099                                                           & 0.078 & 0.074 &  0.072 & 0.068 \\ \hline
\end{tabular}
{\caption{Simulated Ups And d averaged albedo values in the Roman filters as a function of orbital phase for the $T_{int}=0$ case with $0.1$ and $1$ $\mu$$m$ H$_{2}$O cloud particle sizes. Albedo values in the filter correspond to the spectra given in Figure \ref{fig:UpsAnddpphasecolors}.} \label{tab:UpsAnddfilteralbedo}}
\end{table}

 \begin{figure}[t]
  \centering
  \includegraphics[scale=0.60]{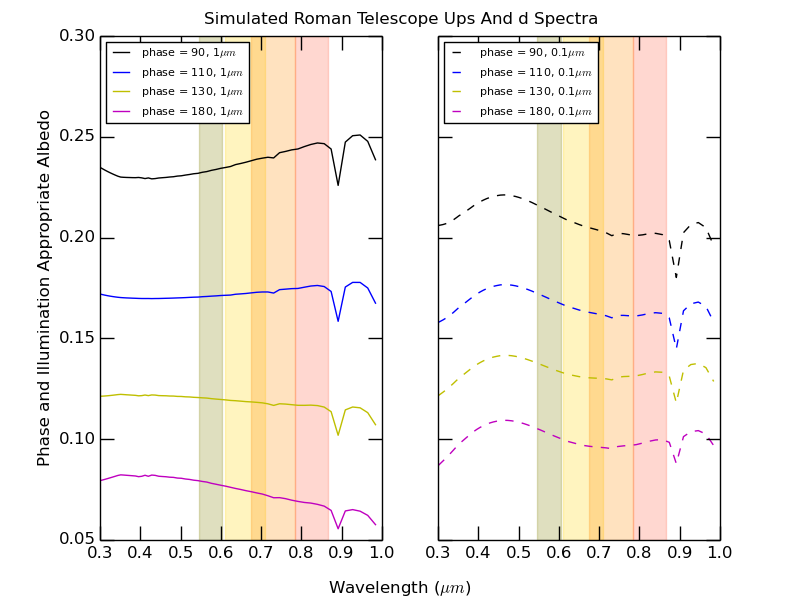}
  \caption{Simulated phase and illumination appropriate Roman model spectra for $\upsilon$ And d at different phases from periastron, as given in the config file included in Section \ref{sec:configfiles} of the appendix.  The albedo shown is the left hand value of Equation 3 in \textcite{2010ApJ...724..189C}, which includes the phase function. The panel on the left shows how spectra varies with phase in the most observationally favorable portions of the orbit for an atmosphere where the water cloud particle size is $1$ $\mu$m. The panel on the right shows similar simulations of spectra, except that the water cloud particle size used was $0.1$ $\mu$m.  Roman band averaged albedos and color-color ratios are given in Tables \ref{tab:UpsAnddfilteralbedo} and \ref{tab:UpsAnddfiltercolor}.}
  \label{fig:UpsAnddpphasecolors}
\end{figure}

\begin{figure*}[t]
  \centering
  \includegraphics[scale=0.70]{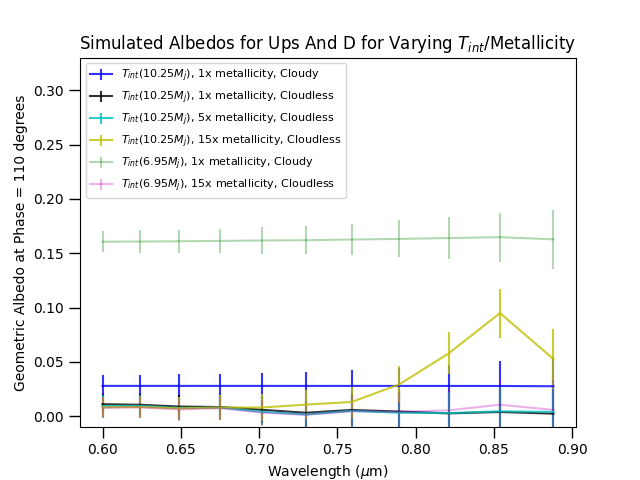}
  \caption{Simulated albedo spectra for $\upsilon$ And d for a variety of metallicity and $T_{int}$ values. Spectra are produced using the observational parameters given in Table \ref{tab:upsanddparam} and a phase of 110 degrees from periastron, which corresponds to one of the most advantageous illumination phases for the planet. The majority of the spectra plotted are for the case where $T_{int}$ corresponds to a planet mass of 10.25 $M_{j}$, but we also plot a case for the lower bound mass of 6.95 $M_{j}$ (the upper bound is very close to the 10.25 $M_{j}$ values). Error bars are 1-sigma values at R=25 for a 400 hour observation; based on those, Roman observations may be able to put bounds on the interior temperature and possibly the metallicity of the planet.  Cases with tholin hazes were also run but not included since transmissivity was $\sim$$>$99.9\% for the haze in this spectral region.}  
  \label{fig:UpsAnddVaryTint}
\end{figure*}

The most observationally favorable portion of the orbit occurs during the planet's approach to apastron, which is why we typically use the apastron-specific atmosphere models in the simulations examining the impact of phase and illumination.  The mean separation cases plotted using geometric albedo in Figure \ref{fig:UpsAndd5xmetalcases} are also potentially a useful model that can be examined given potential latency as the planet moves from periastron towards apastron during the observationally optimal portion of its orbit. However, while the optimal observations may be a little past mean separation, the two cases differ by such a small enough amount that it does not substantially affect our conclusions.  We also restrict our simulations in this case to the haze-free scenarios, given the relatively uncertain nature of potential haze existence and properties. Using PSG and the config file included in Section \ref{sec:configfiles} of the appendix, we simulate the different coronagraph instrument scenes (both the SPC and HLC modes ~\citep{Mennesson2020arXiv}) for our atmosphere model, the appropriate orbital and viewing configuration and the relevant instrument parameters.  This includes the current coronagraph throughput as a function of separation as well as the other optical loss factors and detector noise characteristics.  The phase- and illumination-specific albedo spectra are given in Figures~\ref{fig:UpsAnddpphasecolors} (for the $T_{int}=0$ case) and \ref{fig:UpsAnddVaryTint} (for the cases where $T_{int}$ was modeled based upon \textcite{marley_mark_2018_1309035}).  There are a number of key takeaways from these simulations that are relevant to potential means of extracting atmospheric properties.  

\begin{table}[t]
\begin{tabular}{|c|c|c|c?c|c|c|}
\hline
\multicolumn{4}{|c?}{\textbf{Color/Color Ratio ($\bm{0.1 \: \mu \text{m} \: \text{Particle Size}}$)}}                                                                                            &  \multicolumn{3}{c|}{\textbf{Color/Color Ratio ($\bm{1 \: \mu \text{m}}$)}}                                                                                                                                       \\ \hline 
\textbf{Degrees from}  & \textit{660 nm}/ & \textit{730 nm}/ & \textit{825 nm}/ & \textit{660 nm}/ & \textit{730 nm}/ & \textit{825 nm}/ \\
\textbf{Periastron}  & \textit{575 nm} & \textit{575 nm} & \textit{575 nm} & \textit{575 nm} & \textit{575 nm} & \textit{575 nm}
\\ \hline
\textit{90}                                                                                                                           & \textcolor{Vermillion}{0.968}                                                             & \textcolor{Vermillion}{0.951}                                                             & \textcolor{Vermillion}{0.947}    & \textcolor{SkyBlue}{1.016}  & \textcolor{SkyBlue}{1.031} & \textcolor{SkyBlue}{1.054}                                                     \\ \hline
\textit{110}                                                                                                                           & \textcolor{Vermillion}{0.966}                                                            & \textcolor{Vermillion}{0.952}                                                           & \textcolor{Vermillion}{0.956}      & \textcolor{SkyBlue}{1.007} & \textcolor{SkyBlue}{1.014} & \textcolor{SkyBlue}{1.028}                                                       \\ \hline
\textit{130}                                                                                                                           & \textcolor{Vermillion}{0.969}                                                            & \textcolor{Vermillion}{0.963}                                                            & \textcolor{Vermillion}{0.981}                                                        &  \textcolor{Vermillion}{0.989} &  \textcolor{Vermillion}{0.979} & \textcolor{Vermillion}{0.971}   \\ \hline
\textit{180}                                                                                                                           & \textcolor{Vermillion}{0.949}                                                            & \textcolor{Vermillion}{0.937}                                                            & \textcolor{Vermillion}{0.963}                                                           &  \textcolor{Vermillion}{0.956} &  \textcolor{Vermillion}{0.918} & \textcolor{Vermillion}{0.872} \\ \hline
\end{tabular}
{\caption{Color-color ratios of simulated averaged albedo values for Ups And d in the Roman filters as a function of orbital phase for the $T_{int}=0$ case with $0.1$ and $1$ $\mu$$m$ H$_{2}$O cloud particle sizes. Colors are normalized to the average albedo values for simulations of Ups And d in the 575 nm filter at a particular phase.  Ratios below one (bluer spectra) are colored in sky blue while ratios above one (redder spectra) are in vermillion.  Note that the trend in spectra as a function of phase stays relatively consistent (slightly blue) for the 0.1 $\mu$$m$ cases with increasing phase while the 1 $\mu$$m$ case \textit{flips from redder to bluer spectra} with increasing phase. }} \label{tab:UpsAnddfiltercolor}
\end{table}

For the phase- and illumination-specific albedo spectra for the $T_{int}=0$ case, the 1- and 0.1-$\mu$m cloud particle size cases exhibit different continuum flux values in their spectra at the same phase (see Table \ref{tab:UpsAnddfilteralbedo} for Roman band averaged values).  This is most apparent at the two most extreme phase cases plotted in Figure~\ref{fig:UpsAnddpphasecolors}, where differences between the cases are even apparent in the regions overlaid with the Roman filter bandpasses. This also then relates to another distinguishing pattern between the two cases - that the magnitude of total flux variation as a function of phase for a particle size is also significantly different in the two cases.  As the planet moves from a phase of 90 to 150 degrees, the difference in albedo in some of the Roman filter regions is greater for the 1-$\mu$m case versus the 0.1-$\mu$m case. The Roman averaged band albedos given in Table \ref{tab:UpsAnddfilteralbedo} illustrate this - for example, the 825 nm filter captures the larger variation for models with a 1-$\mu$m cloud particle size for a phase of 90 to 150 degrees from periastron versus the 0.1-$\mu$m particle size case (a difference of 0.178 vs 0.103).

The last prominent, potentially useful characteristic of the simulated spectra in Figure~\ref{fig:UpsAnddpphasecolors} is the potential for a color-color change as the planet moves in its orbit.  The color-color ratios in the Roman bands for these simulations, as normalized to the 575 nm band, are given in table \ref{tab:UpsAnddfiltercolor}.  As an example, for the 1-$\mu$m case, the relative ratio of flux in the bluest Roman filter versus the reddest filter flips as the planet moves from a phase of 90 to 150 degrees (shown by the change in the cell text color from vermillion to sky blue in the 1-$\mu$m cases).  This flip does not occur in the 0.1-$\mu$m case and may be an additional potentially useful diagnostic of the atmosphere depending on the achievable SNR for the observation. While these differences and diagnostics in the spectra are relevant to a change in only the particle size assumed in the model calculations, additional variation of model parameters is likely to produce both degeneracies between potential diagnostics and additional pathways to extracting information about the atmosphere. 

We also examined the phase- and illumination-specific albedo spectra for cases where $T_{int}$ corresponds to a planet mass of 10.25 $M_{j}$ (corresponding to a $T_{eff, mean}$=319K) and 6.95 $M_{j}$ (corresponding to a $T_{eff, mean}$=270K)\citep{marley_mark_2018_1309035}.  The resulting simulated spectra for a selected number of these cases is given in Figure \ref{fig:UpsAnddVaryTint} (a baseline config file used for those cases is given in Section \ref{sec:configfiles} in the appendix).  That figure shows the simulated spectra for a phase of 110 degrees from periastron and in a wavelength region corresponding to the SPC spectroscopy modes. The error bars correspond to 1$\sigma$ shot noise at R=25 for a 400 hour observation.  The 400 hour observation time is based on the exposure time required to achieve previously stated detection limits for Roman \citep{2019arXiv190104050B} and the approximately 2000 hours/3 months of time CGI will nominally receive (including calibration overheads) during its technology demonstration phase \citep{2019arXiv190205569A}. Based on the spectra shown in figure \ref{fig:UpsAnddVaryTint}, it is unlikely that spectroscopic observations would be able to extricate variations in metallicity between 1x and 15x the stellar value for the $T_{eff, mean}$=319K case.  However, for cases with a lower $T_{int}$ and consequently $T_{eff, mean}$, the presence of a water cloud deck that would not exist at the higher temperature cases would be distinguishable.  Thus, observations of $\upsilon$ And d using Roman should be able to put a bound on the value of $T_{int}$ and $T_{eff, mean}$ for the planet.  This is especially true when considering that filter observations would enable shorter observation times and higher signal to noise for the same albedo values. 

\begin{figure*}[]
  \centering
  {\includegraphics[scale=0.50]{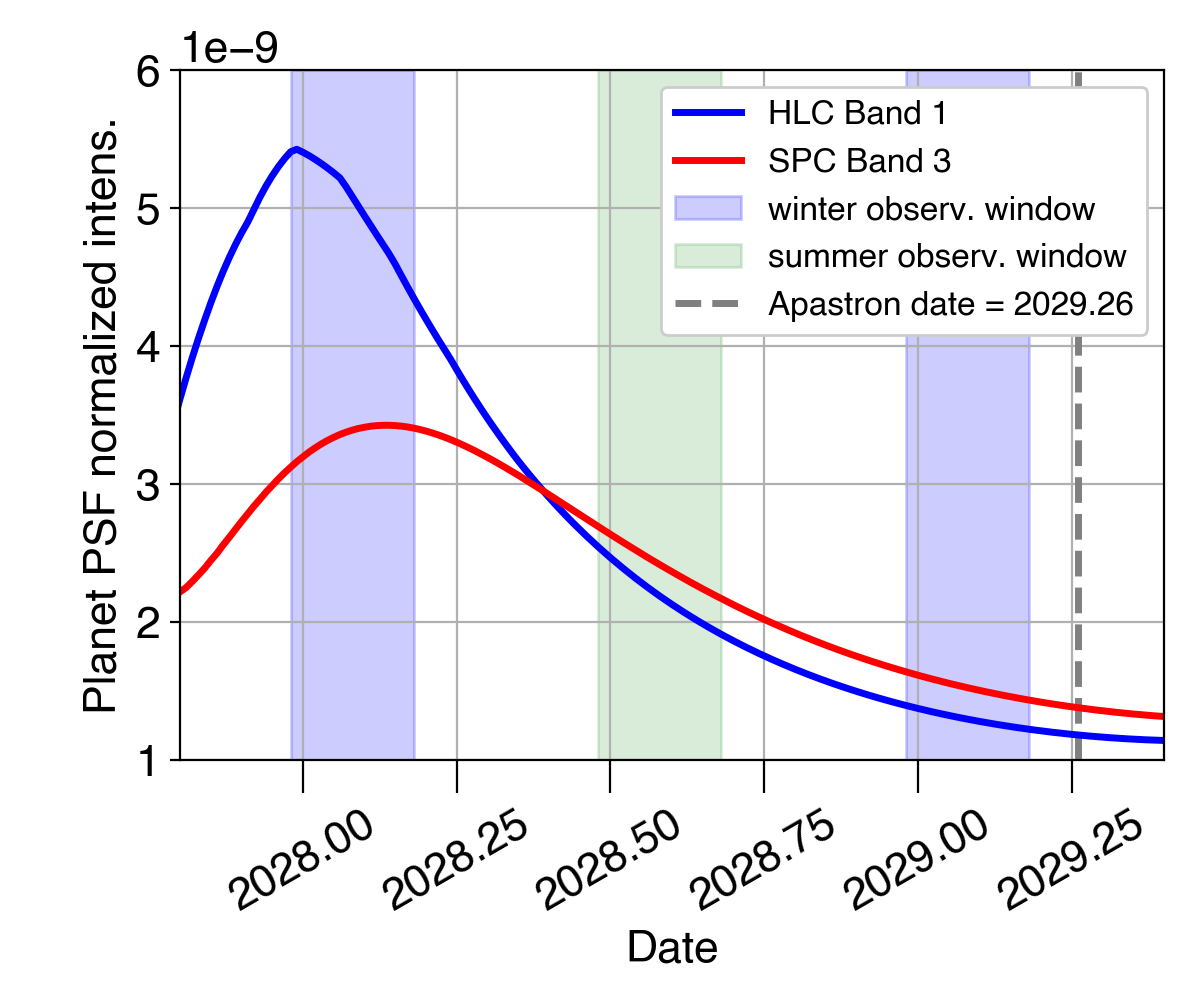}}
  \raisebox{35pt}{\includegraphics[scale=0.30]{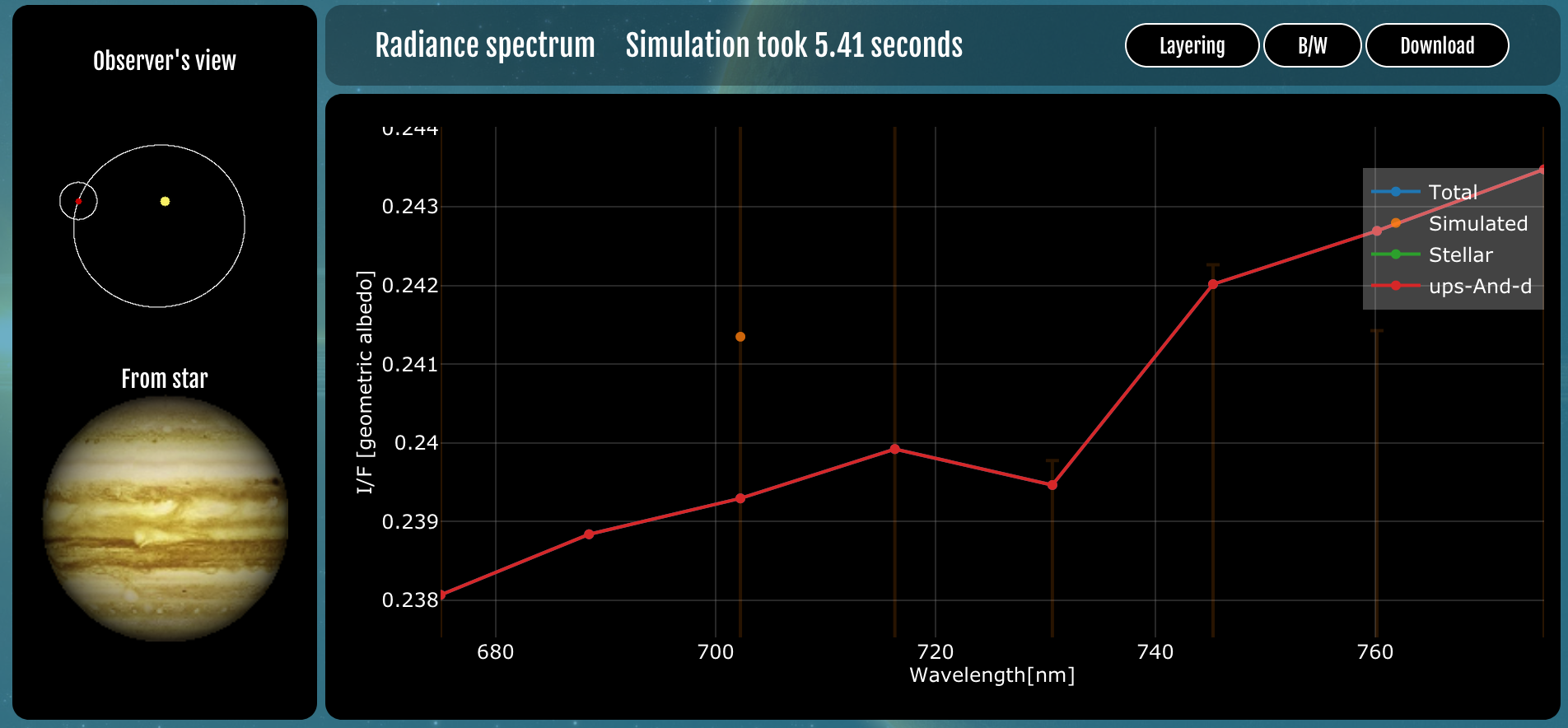}}
  \caption{The left panel illustrates the predicted time-dependent normalized intensity of the point source corresponding to $\upsilon$ And d for two CGI observing modes, the HLC Band 1 (546--604 nm) imaging mode and the SPC Band 3 (675--785 nm) spectroscopy mode. These intensity curves take into account the varying orbital position, phase, and field-dependent coronagraph PSF throughput for the phase- and illumination-specific albedo spectra for the $T_{int}=0$ case with water clouds with 1-$\mu$m cloud particle size. In addition, the shaded rectangles indicate the ranges of dates during which the Roman Space Telescope can point at $\upsilon$ And while meeting the observatory's Sun angle criteria. The right panel shows the simulation result as generated with PSG for the phase=90 case (config file \textbf{for the entire wavelength range included in section \ref{sec:configfiles} of the appendix)}. Error bars are for 400 hours, leading to a SNR=10, and consistent with the realistic Roman CGI simulator.~\citep{Krist2016JATIS, 2019arXiv190104050B, Groff2020AAS} }
  \label{fig:RomanObservables}
\end{figure*}

Finally, with these Roman scene spectra in hand and additional simulated observations for the Roman mission, we examined the optimal time to observe $\upsilon$ And d and its flux during the potential period of the prime Roman mission from 2026--2031 \citep{2019arXiv190205569A}.  The left image of Figure~\ref{fig:RomanObservables} displays the PSF-normalized intensity for $\upsilon$ And d for both the SPC and HLC modes as a function of time and Roman observing windows. $\upsilon$ And d peaks in flux very close to the first winter observing window, exhibiting a normalized intensity of $\sim5\times10^{-9}$ that makes it a viable target for both the Band 3 SPC and Band 1 HLC observing modes. Note that the Band 1 HLC mode has a smaller inner working angle (150 mas, versus 190 mas for the Band 3 SPC), so the higher PSF throughput at the planet's angular separation results in a larger normalized intensity, despite a slightly lower albedo in Band 1 (546--604 nm). The maximum normalized intensity values, $5\times10^{-9}$ for Band 1 HLC (with error $<3\%$) and $3\times10^{-9}$ for Band 3 SPC (with error $5-10\%$ over the spectral range), are within the detection limits of Roman CGI under current best estimates of instrument performance~\citep{2019arXiv190104050B}.
 
We simulated Roman CGI observations of $\upsilon$ And d using two software tools: (i) the Planetary Spectrum Generator, and (ii) a purpose-built Python script for simulating CGI SPC zero-order deviation (ZOD) prism spectroscopy data called \textit{zodprism} \citep{10.1117/12.2562925, 10.1117/12.2563480}. Both tools indicate that $\upsilon$ And d is a viable target for Roman CGI, under the assumptions of orbit parameters and albedo given in Table \ref{tab:upsanddparam}. The right-hand plot of Figure~\ref{fig:RomanObservables} shows one simulated PSG scene assuming a peak intensity observing window, using the config file included in section \ref{sec:configfiles} of the appendix. When a noise model corresponding to the Roman CGI SPC spectroscopy mode is included, PSG predicts reaching a signal-to-noise ratio of 10 in the central wavelength bin of the R=50 spectrum in approximately 400 hours. This time-to-SNR estimate was cross-checked with the instrument team's \textit{zodprism} script, which incorporates a model of the coronagraph point spread function and wavelength-resolved residual starlight pattern~\citep{Krist2016JATIS}, the slit mask and ZOD prism dispersion profile for the spectroscopy mode, the system throughput, and the detector noise parameters~\citep{Groff2020AAS}. The integration time needed to reach the same SNR with the Band 1 HLC imaging mode would be at least $10\times$ less, due to the combination of higher source intensity (Figure~\ref{fig:RomanObservables}) and wider signal bandpass.

\section{Discussion} \label{sec:Discussion}

The viability of $\upsilon$ Andromedae d as a favorable direct imaging (and specifically Roman Space Telescope) target is significant because the planet may be an observationally accessible gas giant that exists in its system's conventional habitable zone.  While planets in colder regions of equilibrium temperature phase space may be thought of as Jupiter and Saturn analogues, $\upsilon$ And d may provide an important exploration of how atmospheric properties vary for planets warmer than the solar system gas giants. These properties, in addition to the planetary bulk properties, may then provide insight into the formation and evolution of $\upsilon$ And d, the $\upsilon$ And A planetary system and gas giants generally. However, in addition to discussion of those topics, it is important to note some additional assumptions and limits in the previous models as well as their ramifications. 

Despite the relatively well-studied nature of the $\upsilon$ And d orbit compared to other RV detected gas giants, the planets' bulk properties are still either unconstrained or possess considerable uncertainty. Chief amongst these is the planetary radius, which has not being observationally constrained and was derived in this paper based upon modeling assumptions and the planetary mass.  While a small discrepancy between the assumed value and the real value is unlikely to change the broader findings of this study, the uncertainty in mass and the lack of measured radius are also important since they inform expections of other properties that may influence the atmosphere, such as the gravity and metallicity.  While the uncertainties in gravity for $\upsilon$ And d is unlikely to significantly change the results of this study, it represents one key variable that would be valuable to further constrain.  This is to say nothing of potentially important second-order effects such as latitudinal flattening, which can cause spatial variations in g by 10s of percent for Saturn \citep{doi:10.1029/2019JE006354}.

The metallicity of the atmosphere, on the other hand, is a fairly important variable that can affect the spectra of the planet significantly in the case of ups And d.  It also can provide important insight into the planetary formation and evolution.  However, an important consideration about inferred or modeled bulk metallicities is that they often rely on the simplistic assumption that metallicity values are uniform across different elements.  In reality, observations of giant planets in the solar system suggest that while broad values of metallicity may be constrained from a group of elements, there can be significant variability from element to element that can be due to unique physical processes or the specific nature of measurement that is taken \citep{2016arXiv160604510A}.  Thus the assumption of a set of metallicity values to test in our spectra simulations is necessarily a contrived parameter meant to elucidate general effects on spectra - the specific numbers are however based on models that link planetary formation and evolution to the atmospheric metallicity values.  The choice of 1-15x the stellar metallicity is based upon the inferred bulk metallicity for Jupiter and Saturn (the two closest solar system giants to $\upsilon$ And d in size and $T_{eq}$ space) and also analyses of exoplanet metal content based upon population mass and radius values (e.g. \textcite{2016ApJ...831...64T}).   The latter study suggests a decreasing relationship in metallicity versus planet mass, with typical values several times the stellar metallicity at the mass range measured for $\upsilon$ And d.  While this assumption again comes with the caveat of the possibility that any individual planet may possess a somewhat unexpected metallicity value (for a variety of factors, such as the efficiency of sequestration of metals to the core), this was used as a guide to the simulations we ran in order to produce potential spectra.  Constraining the metallicity of the atmosphere of $\upsilon$ And d may be an important means of testing hypotheses that suggest there is a connection between the insolation zone of where a planet formed and its bulk metallicity.

In addition to the simplistic assumption for metallicity, a number of other simplifications and assumptions in our models were used, due to the exploratory nature of the work and because of the difficulty in capturing the complexity of giant planet atmospheres.  Giant planet atmospheres observed in our solar system have exhibited complexity in many facets, including but not limited to cloud structure, chemical abundances, atmospheric chemistry processes, vertical and horizontal variability, complicated microphysics and complicated dynamics \citep{2014arXiv1403.4436F}.  Given the complexity of the solar system gas giants and the known effects of certain bulk properties on the atmosphere (for example, the relatively high mass of $\upsilon$ And d may suggest a relatively high internal heat flux for the planet), additional modeling of the planet considering many of these effects is warranted. Additionally, and applicable generally to all potential direct imaging targeted gas giant planets, interpretation of spectra may not directly correspond to or confirm assumptions based on models.  Even in our solar system, many of the extensive NH$_{3}$ and H$_{2}$O cloud decks modeled for the gas giants have not been verified, except in certain regions of strong convective activity.  Perhaps more strikingly for the discussions of cloud deck property effects in our model, the measured altitudes of the main cloud decks in the solar system gas giants do not agree with the predictions of equilibrium cloud condensation theory \citep{2014arXiv1403.4436F}.

The complexity of clouds and other chromophores extends beyond just these bulk morphology properties.  Modeling clouds, even for planets in our solar system and the Earth, is complicated and contains significant uncertainty in composition, microphysics, and radiative properties \citep{2013cctp.book..367M}. Simplifications are necessary when studying exoplanets and may be reasonable given the lower spectral and spatial resolution of relevant observations.  Several effects or considerations which we did not include in our models may warrant more sophisticated study. For example, vertical mixing and transport-induced quenching are important factors in understanding the morphology and composition of clouds \citep{2013cctp.book..367M}, and the relatively   
high value of gravity for $\upsilon$ And d may influence such processes. Another key to understanding and interpreting giant planet atmospheres is understanding the role of hazes, which have significant effect on both Jupiter and Saturn's spectra. Hazes in the upper tropospheres of these planets are dependent on photolytic products that can be produced by ultraviolet radiation. The production of potential hazes on $\upsilon$ And d likely requires modeling of photochemical pathways, analysis of the stellar spectra, shielding effects and a range of other properties for a thorough exploration of the topic.  Our simple tholin-haze case was used to explore the general effect on the spectra of a particular type of haze given that the planet may exist in an flux environment favorable to photolytic haze production relative to Jupiter or Saturn.  Ultraviolet spectra of $\upsilon$ And A taken using Hubble and IUE does exist, and while study of it is beyond the scope of this paper, it may offer insight into the likelihood of hazes in the planet's atmosphere.  However, as a proxy of the relative UV flux environment, we integrated the Lyman-$\alpha$ profile measured \citep{2019ApJ...880..117E} using STIS and compared it to the Sun. $\upsilon$ And A emits more Lyman-$\alpha$ flux than the quiescent Sun \citep{2013ApJ...766...69L} based upon these measurements and consequently the UV environment at $\upsilon$ And d may be significantly more intense than that at Jupiter. 


There are also additional potential influences on the atmosphere and observations of $\upsilon$ And d unique to its orbital and bulk characteristics.  For example, given that both Jupiter and Saturn possess dozens of moons, it may not be unexpected that moons also exist around ups And d (and other gas giants that are targets of direct imaging).  The potential influence on spectra of a simultaneously observed moon or the potential in-fall of material from a volcanically active satellite into the atmosphere may be relevant to observations, particularly in the case of $\upsilon$ And d, where potentially rocky moons would exist in the system's habitable zone.  All of these factors suggest that additional modeling and observations are warranted, given that $\upsilon$ And d may be a promising target for Roman, large ground telescopes that aim to directly image Jupiter-sized planets, and eventually larger space-based telescopes in the future.  


\begin{acknowledgements}

Acknowledgements: We thank Ryan MacDonald and Mark Marley for their help in validation of reflected light spectra simulations with PSG.  We thank Allison Youngblood and Richard Cosentino for useful conversations that improved the quality of the paper. We thank the anonymous reviewer for helpful comments that greatly improved the quality of this paper. This  work  was  funded  in part by  the  WFIRST CGI  Science Investigation Team contract \#NNG16P27C (PI: Margaret Turnbull) and GSFC’s Sellers Exoplanet Environments Collaboration (SEEC). PS acknowledges support by NASA under award number 80GSFC17M0002. GLV acknowledges support for the development of PSG from the NASA’s Exoplanets Research Program (16-XRP16\_2-0071), NASA’s Emerging Worlds Research Program (15-EW15\_2-0175), NASA’s ExoMars/TGO grant, NASA’s Goddard Fellows Innovation Challenge (GFIC-2017). 

\end{acknowledgements}

%
\newpage
\appendix

\section{Atmospheric Temperature and Pressure Profiles for Non-zero $T_{int}$ Values} \label{sec:TPTint}

In addition to simulations of the Ups And d atmosphere that assumed $T_{int}$=0, we also carried out simulations where $T_{int}$ was set to modeled values from \citep{marley_mark_2018_1309035}.  Those values are based on closest values that correspond to $\upsilon$ And d in the Sonora-Bobcat tables \citep{marley_mark_2018_1309035} with respect to the constrained mass and age of the planet.  The masses used are the reference mass value of 10.25 $M_{j}$ and the lower bound on mass for the planet of 6.95 $M_{j}$ (see table \ref{tab:upsanddparam}).  We do not test the upper bound mass since the value is very similar to the reference value, given that the uncertainty is heavily skewed towards lower mass.  The temperature/pressure profiles indicate the possibility that the planet possess a thin water cloud layer for only a small portion of it's orbit for the reference mass value of 10.25 $M_{j}$.

\begin{figure*}[h!]
  \centering
  {\includegraphics[scale=0.42]{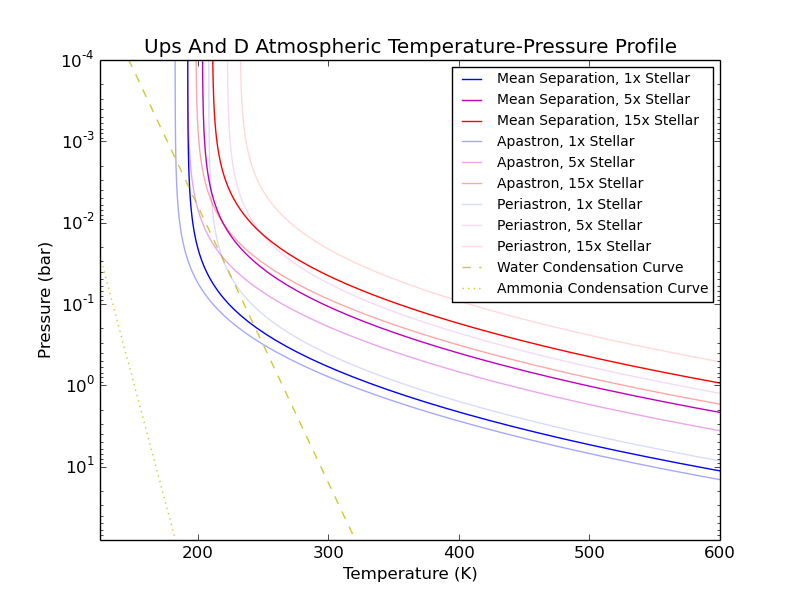}}
  \includegraphics[scale=0.42]{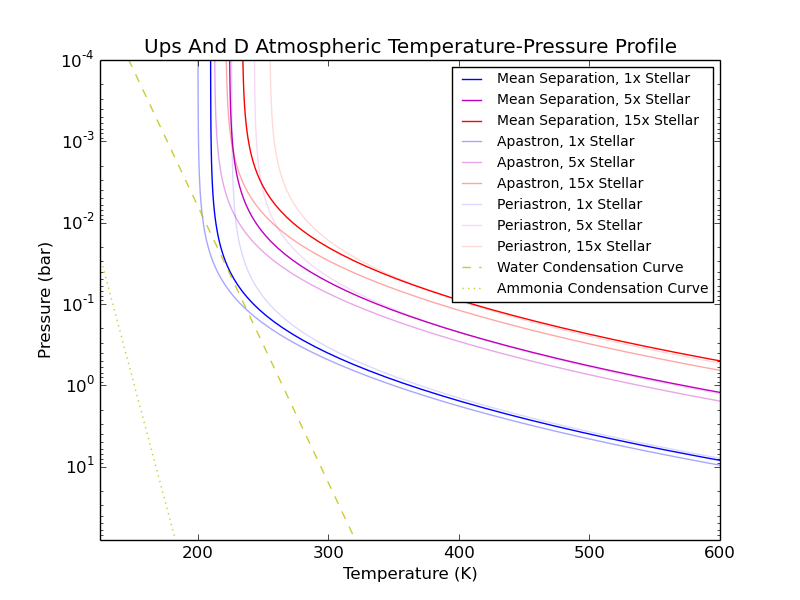}
  \caption{The left panel is a plot of the temperature/pressure profile for $\upsilon$ And d using a value for $T_{int}$ appropriate for the reference mass value of 10.25 $M_{j}$ using \textcite{marley_mark_2018_1309035}. Curves are for the planet at periastron, apastron and mean orbital separation.  The overlaid dashed and dotted lines are water and ammonia condensation curves, respectively, for Jupiter-like gas abundances.  Intersections with the planets atmospheric profile lines indicate potential locations for clouds. The right panel is a similar plot of the temperature/pressure profile for $\upsilon$ And d, but using a value for $T_{int}$ appropriate for the lower bound mass value of 6.95 $M_{j}$.}
  \label{fig:TPTint}
\end{figure*}

\section{Simulation Cloud Properties} \label{sec:cloudprop}

Cloud properties for some of the $\upsilon$ And d simulations run for phase space described in table \ref{tab:upsanddparam}. The total vertical extent of the clouds, the pressure level of the thickest layer of the cloud and the maximum condensate abundance in that layer are given for all the simulations listed in section \ref{sec:UpsAndd} that contained water clouds.

\begin{table}[h!]
\begin{tabular}{|c|c|c|c|c|c|c|}
\hline
\multicolumn{7}{|c|}{\textbf{Cloud Properties of $\upsilon$ And d Simulations}}                                                                                                                              \\ \hline 
\textbf{Cases with} & \textit{Apa, $T_{int}$=0,} & \textit{Apa, $T_{int}$=0,} & \textit{Apa, $T_{int}$=0,} & \textit{Mean, $T_{int}$=0,} & \textit{Mean, $T_{int}$,($6.95  M_{j}$)} & \textit{Apa, $T_{int}$,($10.25  M_{j}$),}  \\
\textbf{Clouds --\textgreater{}} & \textit{1x metallicity} & \textit{5x metallicity} & \textit{15x metallicity} & \textit{5x metallicity} & \textit{1x metallicity} & \textit{1x metallicity} 
\\ \hline
\textit{$P_{max}$ (bars)}                                                                & 0.058                                                            & 0.034                                                             & 0.024                                                            & 0.021  & 0.041 & 0.041                                                     \\ \hline
\textit{$P_{range}$ (bars)}                                                               & 0.79-0.0004                                                            & 0.28-0.0006                                                            & 0.17-0.0009                                                            & 0.098-0.0015     & 0.17-0.0043 & 0.12-0.0085                                                        \\ \hline
\textit{VMR$_{max} (g/g)$}                                                               & 0.017                                                            & 0.075                                                            & 0.13                                                            & 0.20                                                        & 0.015 & 0.015    \\ \hline
\end{tabular}
{\caption{Cloud properties for $\upsilon$ And d simulations included in  section \ref{sec:UpsAndd} that contained water clouds.} \label{tab:clouds}}
\end{table}

\section{Simulation Configuration Files} \label{sec:configfiles}

Below are configuration files that can be uploaded into the web interface or used with the API for PSG in order to simulate spectra for $\upsilon$ And d. Config files for other cases listed in the paper are available upon request.
\newpage

\subsection{Jupiter Reference Config File}
\newpage
\includepdf[pages={1-9}, angle=0]{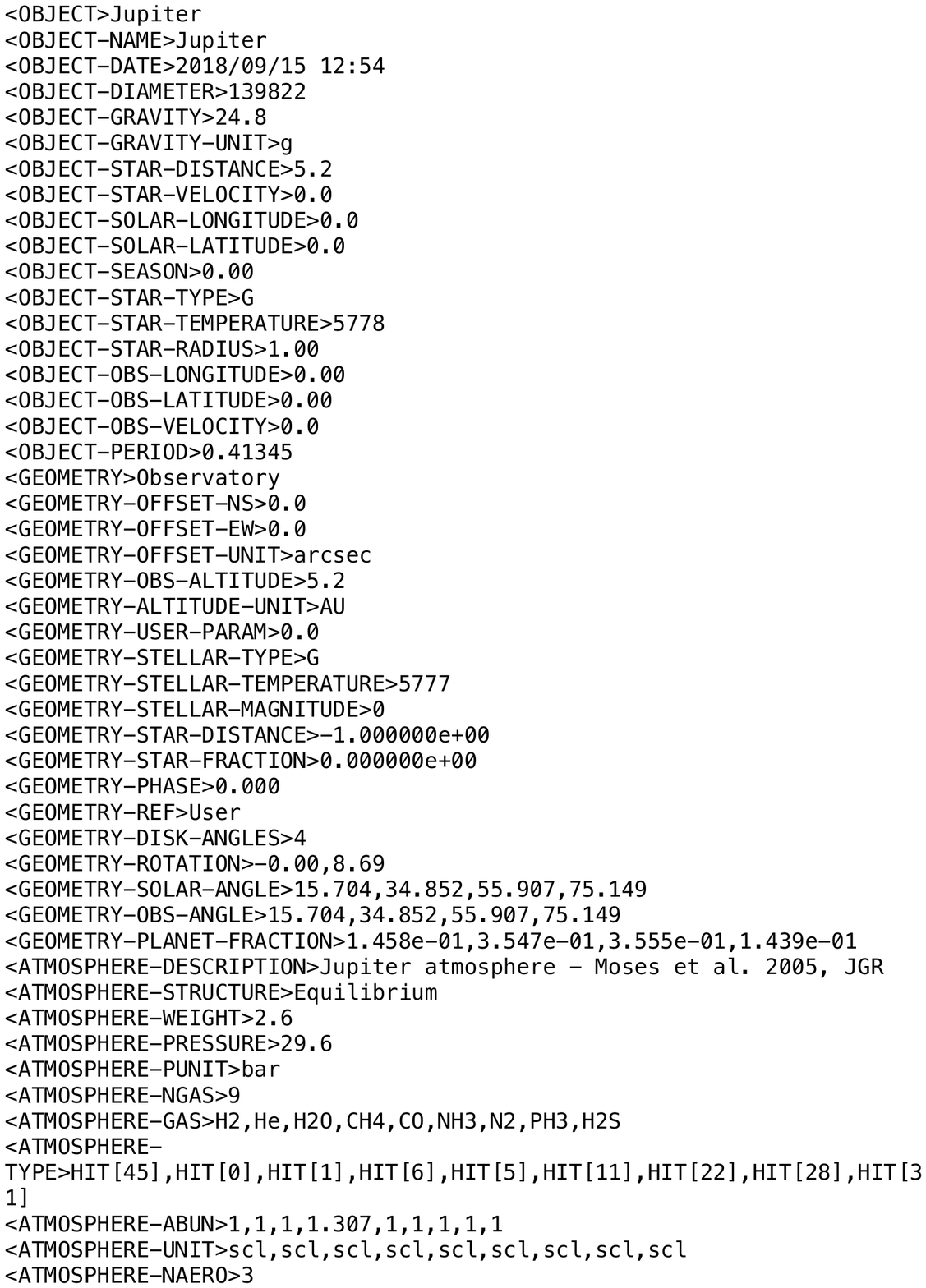}
\newpage
\subsection{$\upsilon$ And d Config File for Interior Temperature = 0 with Noise}
\newpage
\includepdf[pages={1-10}]{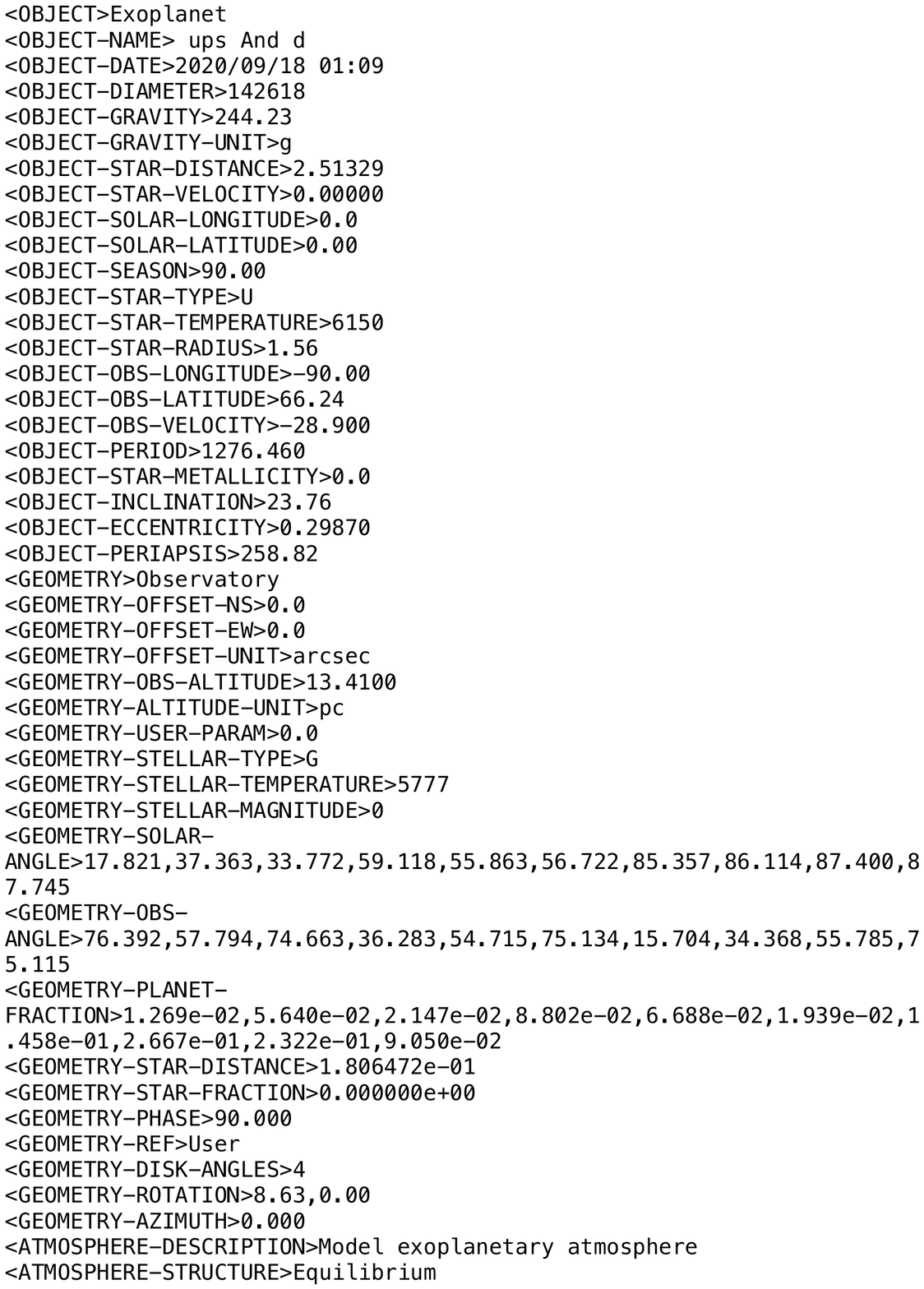}
\newpage
\subsection{$\upsilon$ And d Config File for Interior Temperature = 0 without Noise}
\newpage
\includepdf[pages={1-10}, angle=0]{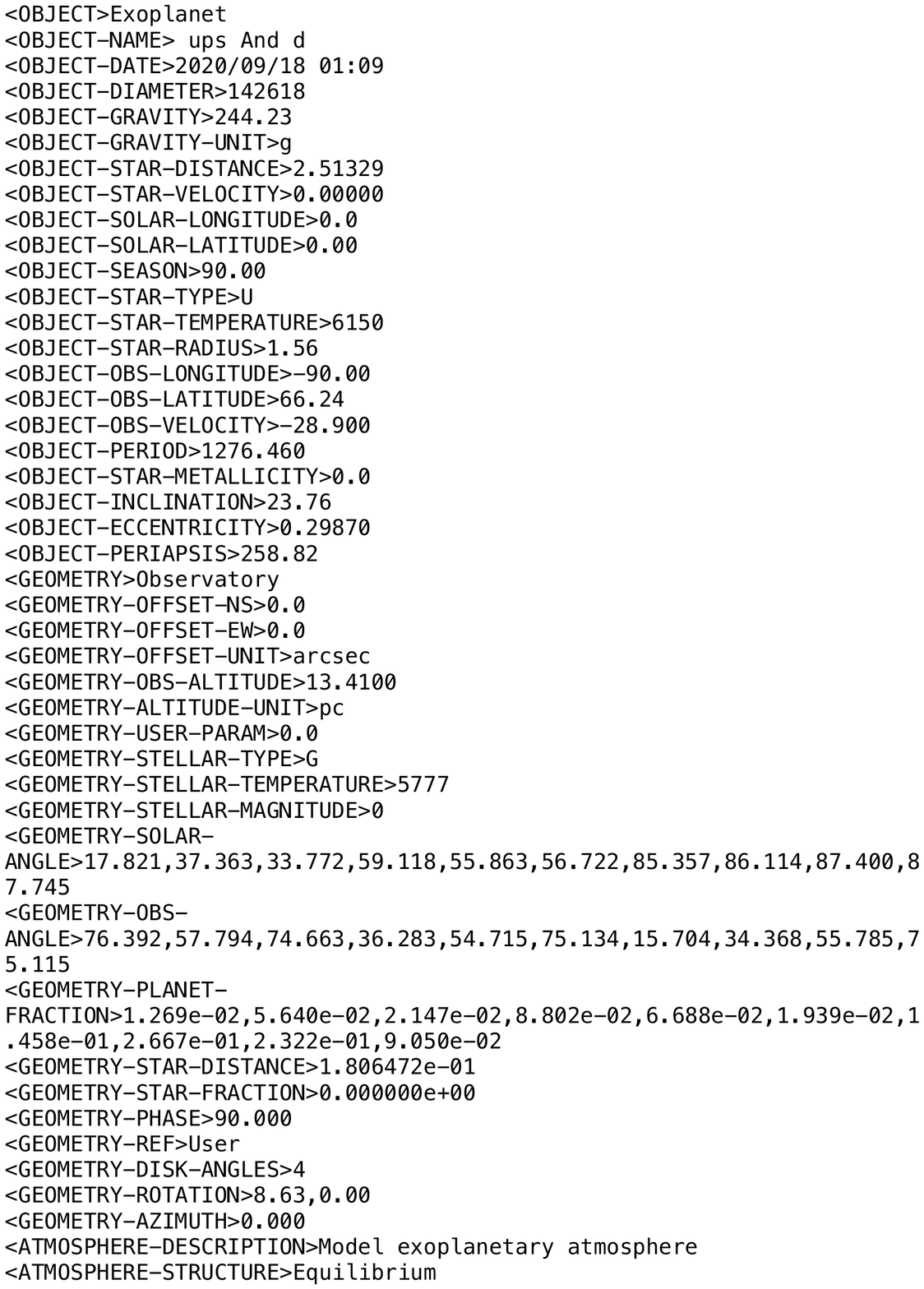}
\newpage





\bibliography{PSGReflectance}{}
\bibliographystyle{aasjournal}



\end{document}